\documentclass[amsmath,amssymb,onecolumn,superscriptaddress]{revtex4}

\usepackage{graphicx}%
\usepackage{dcolumn}%
\def\be{\begin{equation}}
\def\ee{\end{equation}}
\def\beq{\begin{eqnarray}}
\def\eeq{\end{eqnarray}}

\usepackage{bm}
\usepackage{graphicx}
\usepackage{dcolumn}
\usepackage{amsmath}
\usepackage[latin1]{inputenc}
\usepackage{graphicx, psfrag}
\usepackage{amssymb}
\usepackage[colorlinks=true, citecolor=blue, urlcolor = blue, linkcolor= red, bookmarks=true]{hyperref}
\usepackage{float}
\usepackage{amsmath}
\usepackage{amsfonts}
\usepackage{dcolumn}
\usepackage{hyperref}
\usepackage{subfigure}
\usepackage{pgfplots}
\usepackage{epstopdf}
\usepackage{booktabs}
\usepackage{orcidlink}
\usepackage{xcolor}

\begin{document}
\title{Charged strange star model in Tolman-Kuchowicz spacetime in the background of 5D Einstein-Maxwell-Gauss-Bonnet gravity}

\author{Pramit Rej\orcidlink{0000-0001-5359-0655}}
\email[Email:]{pramitrej@gmail.com, pramitr@sccollegednk.ac.in }
 \affiliation{Department of Mathematics, Sarat Centenary College, Dhaniakhali, Hooghly, West Bengal 712 302, India}

\author{Abdelghani Errehymy\orcidlink{0000-0002-0253-3578}\footnote{Corresponding author}}%
\email[Email:]{abdelghani.errehymy@gmail.com}
\affiliation{Astrophysics Research Centre, School of Mathematics, Statistics and Computer Science, University of KwaZulu-Natal, Private Bag X54001, Durban 4000, South Africa}
\affiliation{Laboratory of High Energy Physics and Condensed Matter, Department of Physics, Faculty of Sciences A\"{i}n Chock, Hassan II University of Casablanca, B.P. 5366 Maarif, Casablanca 20100, Morocco}

\author{Mohammed Daoud\orcidlink{0000-0002-8494-2796}}%
\email[Email:]{m$_{}$daoud@hotmail.com}
\affiliation{Department of Physics, Faculty of Sciences, Ibn Tofail University, P.O. Box 133, Kenitra 14000, Morocco}
\affiliation{Abdus Salam International Centre for Theoretical Physics, Miramare, Trieste 34151, Italy}

\begin{abstract}\noindent

In this article, we provide a new model of static charged anisotropic fluid sphere made of a charged perfect fluid in the context of 5D Einstein-Maxwell-Gauss-Bonnet (EMGB) gravity theory. To generate exact solutions of the EMGB field equations, we utilize the well-behaved Tolman-Kuchowicz (TK) {\it ansatz} together with a linear equation of state (EoS) of the form $p_r=\beta \rho-\gamma$, (where $\beta$ and $\gamma$ are constants). Here the exterior
space-time is described by the EGB Schwarzschild metric. The Gauss-Bonnet Lagrangian term $\mathcal{L}_{GB}$ is coupled with the
Einstein-Hilbert action through the coupling constant $\alpha$. When $\alpha \to 0$, we obtain the general relativity (GR) results. 
Here we present the solution for the compact star candidate EXO 1785-248 with mass$=(1.3 \pm 0.2)M_{\odot}$; Radius $= 10_{-1}^{+1}$ km. respectively. We analyze the effect of this coupling constant $\alpha$ on the principal characteristics of our model, such as energy density,  pressure components, anisotropy factor, sound speed etc. We compare these results with corresponding GR results. Moreover, we studied the hydrostatic equilibrium of the stellar system by using a modified Tolman-Oppenheimer-Volkoff (TOV) equation and the dynamical stability through the critical value of the radial adiabatic index.The mass-radius relationship is also established to determine the compactness factor and surface redshift of our model.
In this way, the stellar model obtained here is found to satisfy the elementary physical requirements necessary for a physically viable stellar object. 

\end{abstract}

\maketitle
\section{Introduction}\label{sec1}
Owing to the difficulties encountered by the general theory of relativity (GTR) in explaining the anomalous behavior of gravitational events such as the accelerated expansion of the cosmos in a late-time \cite{Riess-Perlmutter:1998,Riess-Perlmutter:1999}, alternative or extended gravity theories have suddenly gained considerable importance. Conjecturing the presence of exotic matter fields, including quintessence fields (QFs), ghost fields (GFs), dark energy (DE), and dark matter (DM), to name a few, is one approach to solving this problem. There is currently no empirical evidence for these conjectures, but a variety of experiments are being carried out. In this concern, de Rham \cite{de-Rham:2014} proposes that the graviton is not massless but actually bears a small mass to explain the dark sector. This has several implications for physics, which have already been addressed previously in the literature. Reexamining the geometrical side of the field equations offers an alternative approach, meanwhile higher curvature impacts may have a role to play. Specifically, the Einstein-Gauss-Bonnet (EGB) theory has shown to be promising in this aspect and is hence widely investigated. It should be noted that the EGB is part of a more generic category of theories named Lovelock's polynomial Lagrangians which are the most comprehensive tensor theory yielding at most 2nd-order motion equations. The most common theory is owed to Horndeski \cite{Horndeski:1974} if one allows the Lagrangian to involve both tensor and scalar fields. The reality that the Gauss-Bonnet Lagrangian naturally manifests in the applicability of heterotic string theory at the low energy limit \cite{Gross:1999} provides another compelling argument in favor of the EGB theory. For inhomogeneous distributions of dust \cite{Hansraj:2022,Mustafa:2022} and null dust \cite{Ghosh:2014}, the causal structure of the singularities deviates from general relativity. The challenge we are studying seems to be whether gravitational effects with higher curvature play an important role in stellar evolution. In fact, Einstein's GTR forms the basis for the majority of our knowledge of discoveries in relativistic astrophysics. So it is therefore quite natural to wonder what consequence the higher curvature contributions have on stellar configurations if the GTR is to be replaced by a higher curvature theory that maintains 2nd order motion equations and shrinks to GTR in the solar system scale constraint.

Although 5-dimensional celestial bodies are not physically obtainable, their occurrence has not been excluded. Kaluza \cite{Kaluza:1921} and Klein \cite{Klein:1926} made the first advances in higher dimensional gravity by analyzing a 5-dimensional manifold and explaining the behavioral patterns of the electrodynamic field by four components of the metric tensor, ten by the typical 4-dimensional space-time manifold, and one additional dimension by a scalar field. Following that, modern works on brane-world cosmologies requiring higher dimensions have received significant coverage, despite the fact that its emphasis has since waned. The additional dimensions are usually explained by the fact that they are topologically coiled and extremely tiny in size. Keep in mind that the LHC experiment probed for additional dimensions on a large scale but was unsuccessful in finding any. Nevertheless, this does not exclude the possibility of additional dimensions at the microscopic level. Indeed, quantum field theory actually requires space-time dimensions of the size of Ten and Eleven. As we demonstrate in this paper, despite their small size, they may have a significant impact on certain features of the gravitational field.

For a long time, the existence of spherically symmetric static black hole (BH) solutions in the EGB theory paradigm has been studied extensively \cite{Tangherlini:1963,Myers:1986,Myers:1988}. Subsequently, several other factors have been explored for the GB solution in de Sitter (dS) and anti-de Sitter (AdS) space \cite{Cai:2002,Cai:2004}, including thermophysical properties related to the cosmological horizon and the BH horizon. Additionally, some authors have recently examined specific solutions relating to BHs in great detail (see e.eg. Refs. \cite{Giacomini:2015,Xu:2015,Aranguiz:2016,Ghosh:2017}). Considerable efforts have been made to study the geodesic motion of a test particle \cite{Bhawal:1990}, the radius of photon spheres \cite{Gallo:2015}, the Hawking evaporation of AdS BHs \cite{Wu:2021}, the phase transition of RN-AdS BHs \cite{Xu:2019}, solutions of regular BHs \cite{Ghosh:2018}, solutions of wormholes fulfilling energy conditions \cite{Maeda:2008,Mehdizadeh:2015}, and the gravitational collapse of an incoherent spherical dust cloud was proposed in \cite{Jhingan:2010,Maeda:2006,Zhou:2015,Abbas:2015}.

The study of EGB gravity may also be important with respect to the prospect of overcoming some issues in strong field regimes. The study of cosmic structures may also offer important restrictions on the alternative theories of gravity that are being considered in this context. A significant number of mass-radius relations are accessible at the moment through electromagnetic measurements of extreme events like short gamma-ray bursts (SGRBs), and more recently through gravitational waves (GWs). It is worth noting that the theoretical building of the neutron star (NS) equation of state (EoS) is strongly constrained by the observational data for NSs with masses around $2~M_{\odot}$ \cite{Demorest:2010,Antoniadis:2013}. Until now, the composition and structural properties of relativistic compact bodies are not well recognized in detail. Many solutions to Einstein's gravitational field equations defining the internal construction of relativistic compact bodies have now been derived, all assuming static and spherical symmetry. Due to the complexity of the Einstein field equations, however, it is typically difficult to find exact solutions in GR that are physically realizable. In light of modified theories of gravity, the situation gets more challenging. So, in order to arrive at exact solutions, researchers employ a number of mathematical methodologies. Many such concepts have been discussed, in particular, an algorithm that produces all regular static, spherically symmetric perfect-fluid solutions of Einstein's equations by selecting a single monotone function \cite{Lake:2003} and its enlargement to locally anisotropic fluids in \cite{Herrera:2008}. With compact objects, however, there is flexibility in selecting the appropriate interior solutions. Schwarzschild found the first exact solution to Einstein's field equations in 1916 \cite{Schwarzschild:1916}, which improved our ability to predict many physical events. By using spherically symmetric perfect fluid solutions of Einstein's equations, Tolman \cite{Tolman:1939} initially derived the simplest model to represent stellar interiors. Following that, several works--both theoretical and observational--on various aspects of modified gravity theories have been reported in recent literature \cite{refa0,refa1,refa2,refb0,refb1,refb2,refb3,refb4,refb5,refc0,refc1,refc2,refc3,refc4,refc5,refd0,refd1,refd2,refd3,refd4,refe0,refe1,reff0,reff1,reff2,reff3,reff4,reff5}.

The foregoing arguments encouraged us to make a thorough study of the stability and appropriateness of anisotropic solutions for relativistic compact stars. In this article, we are mainly interested in exploring the salient features of the static-charged anisotropic fluid sphere consisting of a charged perfect fluid in the framework of 5D EMGB gravity theory. We are using the well-behaved TK {\it ansatz} together with a linear EoS of the form $p_r=\beta \rho-\gamma$, (where $\beta$ and $\gamma$ are constants) to generate exact solutions of the EMGB field equations and analyze the effect of the Gauss-Bonnet Lagrangian term $\mathcal{L}_{GB}$ which is coupled with the Einstein-Hilbert action through the coupling constant $\alpha$ on the principal physical characteristics of the stellar model. We would also like to point out that the characteristics of the matter configuration and higher dimensional effects are directly related.

The manuscript is organized in the following way: In Sects. \ref{Sec2}, \ref{Sec3} and \ref{Sec4}, we are providing the relevant EGB gravity formalism as well as the EMGB field equations and their solutions, respectively. The discussion of matching conditions for constant evaluations can be found in Sect. \ref{Sec5}. In Sect. \ref{Sec6}, we have been discussing significant results for the actual charged model and its stability in Sect. \ref{Sec7}. Finally, we are presenting our concluding remarks in Sect. \ref{Sec8}. Throughout the paper, we have employed the essentially positive signature ($-,~+,~+,~+,~+$).

\section{Einstein-Gauss-Bonnet gravity}\label{Sec2}
To construct the field equations in EGB gravity, we need a modified action that differs from the Einstein scenario. In this study, we will be working on five dimensions. In five-dimensional spacetime with a matter field, the modified action for EGB gravity can be written as,
\begin{eqnarray}
 \label{action}
	\mathcal{I}_{G}=\frac{1}{16 \pi }\int d^{5}x\sqrt{-g}\left[ R-2\Lambda +\alpha \mathcal{L}_{\text{GB}} \right]+
\mathcal{S}_{\text{matter}}.
\end{eqnarray}
where $g$ signifies the determinant of the metric $g_{ij}$, $R$ displays the five-dimensional Ricci scalar, and $\Lambda$ stands for the cosmological constant, whereas $\mathcal{S}_{\text{matter}}$ denotes the action related to the matter field. The GB coupling constant $\alpha$ is correlated to the string tension in string theory and has length squared dimensions, representing ultraviolet corrections to Einstein's theory. It should be noted that according to Maedaa \cite{Maeda:2006}, the coupling constant $\alpha$ must be nonnegative for Minkowski spacetime to be stable. Consequently, we will constrain our study to the case where the GB coupling constant $\alpha$ is non-negative. The Ricci scalar, Ricci tensor, and Riemann curvatures are specifically combined in the GB term $\mathcal{L}_{\text{GB}}$, which is stated by 
\begin{equation}
\mathcal{L}_{\text{GB}}=R^{ijkl} R_{ijkl}- 4 R^{ij}R_{ij}+ R^2\label{GB}.
\end{equation}
We also incorporate the matter action $\mathcal{S}_{\text{matter}}$, which triggers the stress-energy tensor for the matter field, ${T}_{ij}$. We now get the following equations of motion by varying the action (\ref{action}) in relation to the metric $g_{ij}$, 
\begin{equation}\label{eq3}
G_{ij}+\alpha H_{ij} = \frac{8 \pi G}{c^4}  T_{ij},~~\text{where}~~{T}_{ij}=-\frac{2}{\sqrt{-g}}\frac{\delta\left(\sqrt{-g}\mathcal{S}_m\right)}{\delta g^{ij}}. 
\end{equation}
Here, the Einstein tensor and the tensor containing the contributions from the GB component are denoted by the symbols $G_{ij}$ and $H_{ij}$, respectively. These both tensors are explicitly expressed as 
\begin{eqnarray}
&&\hspace{-0.7cm}  G_{ij} = R_{ij}-\frac{1}{2}R~ g_{ij},\nonumber\\
&&\hspace{-0.7cm} H_{ij} =  2\Big( R R_{ij}-2R_{ik} {R}^k_j -2 R_{ijkl}{R}^{kl} - R_{ikl\delta}{R}^{kl\delta}_j\Big) - \frac{1}{2}~g_{ij}~\mathcal{L}_{\text{GB}}.~~\label{FieldEq}
\end{eqnarray} 
The GB term does not exist for $n \leq 4$ but actively contributes to the curvature of the spacetime for $n \geq 5$ since it is a topological invariant in the $4$-dimensional spacetime. In writing the Einstein-Maxwell-Gauss-Bonnet (EMGB) field equations, here we are using geometrized units, and thus we have taken $G = c = 1$ throughout the discussion.
\section{Field Equations}\label{Sec3}
\label{sec:3}The $5$-dimensional line element for a static spherically symmetric spacetime has the standard form
\begin{eqnarray}
\label{5} ds^{2}& =& -e^{2\nu(r)} dt^{2} + e^{2\lambda(r)} dr^{2} +
 r^{2}(d\theta^{2} + \sin^{2}{\theta} d\phi^2 +\sin^{2}{\theta} \sin^{2}{\phi^2}
 d\psi),
\end{eqnarray}
in coordinates ($x^i = t,r,\theta,\phi,\psi$). By considering the comoving fluid velocity as $u^a=e^{-\nu}\delta_0^a$, the Einstein-Maxwell-Gauss-Bonnet (EMGB) field equation (\ref{eq3}) yields the following set of independent equations in view of the metric (\ref{5}),
\begin{eqnarray}
\label{7a}\kappa \rho + E^2&=& -\frac{3}{e^{4\lambda }r^3} \Bigg[4\alpha \lambda'
+re^{2\lambda}-re^{4\lambda}- r^2 e^{2\lambda}\lambda'  -4\alpha e^{2\lambda}\lambda'\Bigg],\label{fe1}\\
\label{7b} \kappa p_r - E^2 & = &\frac{3}{e^{4\lambda }r^3}
\Big[-re^{4\lambda}+ \Big(r^2 \nu' +r +4\alpha \nu'\Big)e^{2\lambda} -4\alpha \nu'\Big] , \label{fe2}\\
\label{7c} \kappa p_t + E^2&=& \frac{1}{e^{4\lambda }r^2} \Big[- e^{4\lambda
}- 4\alpha \nu''+ 12 \alpha \nu' \lambda' -4 \alpha
(\nu')^2\Big]
 +\frac{1}{e^{2\lambda }r^2} \Big[1- r^2 \nu' \lambda' +2r \nu'
-2r  \lambda' +r^2(\nu')^2 \Big] \nonumber \\
&& +\frac{1}{e^{2\lambda }r^2} \Big[r^2 \nu'' -4\alpha
\nu'\lambda' + 4\alpha (\nu')^2  +4\alpha \nu''\Big],\label{fe3}\\
\label{7d} \kappa \sigma &=& \frac{2}{r^2} e^{-\lambda /2} (r^2 E)',\label{fe4}
\end{eqnarray}
where $\rho$, $p_r$ and $p_t$ respectively denote the matter density, radial and transverse pressure of the fluid. Here $E$ is the electric field intensity, $\sigma$ is the proper charge density and $\kappa= 8\pi$. Note that a $\prime$ denotes the differentiation with respect to the radial coordinate $r$. 
If $q(r)$ represents the total charge contained within the
$5-$ D sphere of radius $r$, then it can be defined by the relativistic Gauss law as
\begin{equation}
q(r) = 2 \pi^2 \int_0^r \sigma r^3 e^{\lambda} dr.  \label{charge}
\end{equation}
\section{Solution of the Field Equations}\label{Sec4}
To solve the above field equations (\ref{fe1})-(\ref{fe3}) we utilize the following {\em ansatz}
\begin{eqnarray}\label{elambda}
e^{\lambda (r)}= 1 + ar^2 + br^4, e^{\nu (r)} &=& C^2 e^{B r^2},\label{tk}
\end{eqnarray}
where $a, b, B, C$ are constants. These metric potentials conform to the well-known Tolman-Kuchowicz \cite{tol, kuch} spacetime.

Using this Tolman-Kuchowicz ansatz given in Equations (\ref{tk}), the field equations (\ref{fe1})-(\ref{fe3}) can be rewritten as,
\begin{eqnarray}\kappa \rho + E^2&=& \frac{3}{r^3 \Psi^4}\Big[-\frac{8 \alpha r (a + 2 b r^2)}{\Psi} +
   8 \alpha r (a + 2 b r^2) \Psi +    2 r^3 (a + 2 b r^2) \Psi - r \Psi^2 +
    r \Psi^4)\Big]\label{fe4}\\
    \kappa p_r - E^2&=&\frac{3}{r^3 \Psi^4}\Big[-8 \alpha B r + (r + 8 \alpha B r + 2 B r^3) \Psi^2 -
    r \Psi^4\Big]
    \label{fe5}\\
    \kappa p_t + E^2&=&\frac{1}{\Psi^5}\Big[48 \alpha B (a + 2 b r^2) +
   8 \alpha B (a - 2 B + b r^2) \Psi -
   4 \Big(a - 2 B + 2 a \alpha B + A_1 r^2\Big) \Psi^2 - (a + b r^2) \Psi^4 \nonumber\\&&- \Psi^3 \Big(a + 2 B + b r^2 - 4 B^2 (4 \alpha + r^2)\Big)\Big] \label{fe6}
    \end{eqnarray}

The anisotropic factor $\Delta = p_t - p_r$ assumes the following form
\begin{eqnarray}
\Delta&=&\frac{1}{4\pi \Psi^5} \Big[24 \alpha B (a + 2 b r^2) -
   8 \alpha B (a + B + b r^2) \Psi  -
   2 (a - 2 B + 8 a \alpha B + A_2 r^2) \Psi^2 + (a + b r^2) \Psi^4 \nonumber\\&&+ \Psi^3 \Big(a + b r^2 + 2 B (-2 + 4 \alpha B + B r^2)\Big)\Big]-\frac{E^2}{4\pi} \label{del}
\end{eqnarray}

 where,
 \begin{eqnarray}
 \Psi&=&(1 + a r^2 + b r^4),\nonumber\\
A_1&=&-a B + b (2 + 6 \alpha B),\nonumber\\
A_2&=&-a B + 2 b (1 + 6 \alpha B)\nonumber
\end{eqnarray}
Along with TK-metric we also assume that the radial pressure $p_r$ is related to the matter density $\rho$ by a linear
equation of state,
\begin{eqnarray}
p_r &=& \beta \rho - \gamma \label{eos}
\end{eqnarray}
where $\beta$ and $\gamma$ are positive constants.

Adding Eqs. (\ref{fe4})-(\ref{fe5}) we get,
\begin{eqnarray}
\rho + p_r &=& \frac{1}{4\pi \Psi^5} \Big[3 \Big(a + B + (2 b + a B) r^2 + b B r^4\Big) \Big(1 + (4 \alpha + r^2) (a + b r^2) (1 + \Psi)\Big)\Big] \label{pr1a}
\end{eqnarray}
Solving Eqn. (\ref{pr1a}) with the help of Eqn. (\ref{eos}), the matter density($\rho$) and radial pressure($p_r$) are obtained as,
\begin{eqnarray}
\rho &=& \frac{1}{(1 + \beta)} \bigg[\gamma + \frac{1}{4 \pi \Psi^5}\bigg(3 \Big(a + B + (2 b + a B) r^2 +
   b B r^4\Big) \Big(1 + (4 \alpha + r^2) (a + b r^2) (1 + \Psi)\Big)\bigg)\bigg] \label{r1}\\
  p_r &=& \frac{1}{4(1 + \beta)} \bigg[-4\gamma + \frac{1}{\pi \Psi^5}\bigg(3\beta \Big(a + B + (2 b + a B) r^2 + b B r^4\Big)
    \times \Big(1 + (4 \alpha + r^2) (a + b r^2) (1 + \Psi)\Big)\bigg)\bigg] \label{r2}
\end{eqnarray}
By using the expression of $\rho$ in Eqn. (\ref{fe1}) the expression for the electric field intensity is obtained as,
\begin{eqnarray}
E^2 &=&  \frac{3}{r^3 \Psi^4}\Big[-\frac{8 \alpha r (a + 2 b r^2)}{\Psi} +
   8 \alpha r (a + 2 b r^2) \Psi +
   2 r^3 (a + 2 b r^2) \Psi - r \Psi^2 +
    r \Psi^4)\Big] -\frac{8\pi}{(1 + \beta)} \bigg[\gamma + \nonumber\\
    && \frac{1}{4 \pi \Psi^5}\bigg(3 \Big(a + B + (2 b + a B) r^2 +
   b B r^4\Big) \Big(1 + (4 \alpha + r^2) (a + b r^2) (1 + \Psi)\Big)\bigg)\bigg] \label{en}
\end{eqnarray}

Consequently the expression for the transverse pressure $p_t$ is obtained as,
\begin{eqnarray}
p_t &=& \frac{1}{8\pi \Psi^5}\bigg[48 \alpha B (a + 2 b r^2) +
   8 \alpha B (a - 2 B + b r^2) \Psi -
   4 \Big(a - 2 B + 2 a \alpha B + A_1 r^2\Big) \Psi^2 - (a + b r^2) \Psi^4 \nonumber\\&&- \Psi^3 \Big(a + 2 B + b r^2 - 4 B^2 (4 \alpha + r^2)\Big)\bigg] - \frac{E^2}{8\pi}  \label{pr1}
\end{eqnarray}
\section{Exterior spacetime and boundary condition}\label{Sec5}
\textcolor{blue}{In order to determine the constant parameters i.e., $a, b, B,$ and $C$ for our proposed model it is necessary to match the interior space-time solution in a smooth way with an appropriate static and spherically symmetric exterior vacuum Schwarzschild solution. Since here we are investigating the interior anisotropic solution in $5$-D Einstein-Gauss-Bonnet gravity, the most suitable and appropriate static exterior vacuum Schwarzschild solution is proposed by Boulware-Deser \cite{Boulware}. This Einstein-Gauss-Bonnet-Schwarzschild solution is expressed by the following line element:}
 \begin{eqnarray} \label{exterior}
ds^2 &=& -{\cal F }(r)dt^2 + \frac{dr^2}{{\cal F}(r)} + r^{2}\big(d\theta^{2} + \sin^{2}{\theta} d\phi^2
 +\sin^{2}{\theta} \sin^{2}{\phi^2}
 d\psi^2\big)
\end{eqnarray}
where ${\cal {F}}(r) = K + \frac{r^2}{4\alpha}\left(1 - \sqrt{1 + \frac{8\alpha M}{r^4} - \frac{8\alpha Q^2}{3r^6}}\right)$. The quantities $M$ is the gravitational mass, $Q$ represents the charge of the fluid as determined by an isolated observer at spatial infinity and $K$ is an arbitrary constant.

Using continuity of the metric functions and their derivatives, namely $g_{rr}$, $g_{tt}$ and $\frac{\partial g_{tt}}{\partial r}$ across the boundary $r=R$ we get,
\begin{eqnarray}
\frac{1}{1 + aR^2 + bR^4} &=& {\cal {F}}(R), \label{ex1}\\
C^2 e^{BR^{2}} &=& {\cal {F}}(R), \label{ex2}\\
2BRC^2 e^{BR^{2}} &=& \frac{-4 \sqrt{3} \alpha Q^2 + 3R^6 \Big(Q_1- \sqrt{3} \Big)}{6 \alpha R^5 Q_1}.\label{ex3}
\end{eqnarray}
where, $Q_1=\sqrt{3 - \frac{8 \alpha (Q^2 - 3 M R^2)}{R^6}}$

Now by solving Eqn. (\ref{ex1}), we obtain the expression for $b$ as a function of $\alpha$ and $a$ as,
\begin{eqnarray}
b &=& \frac{12 \alpha (1 - K - a K R^2) + R^2 (1 + a R^2)(Q_2 -3)}{12 \alpha K R^4 + 3 R^6 - Q_2 R^6}
\end{eqnarray}
where, $Q_2=\sqrt{9 - \frac{24 \alpha (Q^2 - 3 M R^2)}{R^6}}$

Now from the condition $p_r(r = R) = 0$ we get,
\begin{eqnarray}
  0 &=& \frac{1}{4(1 + \beta)} \bigg[-4\gamma + \frac{1}{\pi \Psi_R^5}\bigg(3\beta \Big(a + B + (2 b + a B) R^2 + b B R^4\Big)
    \Big(1 + (4 \alpha + R^2) (a + b R^2) (1 + \Psi_R)\Big)\bigg)\bigg] \label{m1}
\end{eqnarray}
where, $\Psi_R=(1 + a R^2 + b R^4)$
At the center of the star the electric field is obtained as,
\begin{eqnarray}
  E^2(0) &=& \frac{ 6 a (1 - 8 \alpha B + 2 \beta + 8 a \alpha \beta) -6 B - 8 \gamma \pi}{1 + \beta} \label{m2}
\end{eqnarray}
We know from regularity of electric charge that the electric field vanishes at the center of the fluid configuration, $E^2(r=0)=0$.
Thus by simultaneously solving the Eqns. $p_r(r = R) = 0$ and $E^2(r=0)=0$, we obtain the expressions for $B$ and $\gamma$ as functions of $\alpha, \beta$, $a$ and $b$ as,
\begin{widetext}
\begin{eqnarray}
B &=& \Bigg[\Psi_R^4 \bigg(a + 2 a \beta + 8 a^2 \alpha \beta - \frac{1}{\Psi_R^5} \Big[\beta \Big(a + 2 b R^2\Big) \Big(1 + (4 \alpha + R^2) (a + b R^2) (1 + \Psi_R)\Big)\Big]\bigg)\Bigg] \Biggm/ \Bigg[8 a^5 \alpha R^8 + \nonumber\\
   &&(1 + b R^4)^4 +4 a^3 R^4 \Big(1 + b R^4\Big) \Big(R^2 + 12 \alpha (1 + b R^4)\Big) + a^4 R^6 \Big(R^2 + 32 \alpha (1 + b R^4)\Big)+\beta \Big(1 + b R^2 (4 \alpha \nonumber\\
   && + R^2)(2 + b R^4)\Big)+2 a \Big(1 + b R^4\Big) \Big(R^2 \big(\beta + 2 (1 + b R^4)^2\big) +4 \alpha \big(\beta + (1 + b R^4)^3\big)\Big) +a^2 R^2 \Big(R^2 \big(\beta +\nonumber\\
   && 6 (1 + b R^4)^2\big)+4 \alpha \big(\beta + 8 (1 + b R^4)^3\big)\Big)\Bigg], \label{b1}\\
   \gamma &=& \Bigg[3 \beta \bigg(2 a (1 + 4 a \alpha) (1 + \beta) + \Big[2 b + a \Big(a + 16 \alpha b + 2 a (1 + 4 a \alpha) \beta\Big)\Big] R^2 + a b \Big(1 + (2 + 8 a \alpha) \beta\Big) R^4\bigg)\times\nonumber\\&&
   \bigg(1 + (4 \alpha + R^2) (a + b R^2) (1+\Psi_R)\bigg)\Bigg]\Biggm/ \Bigg[4 \pi \Psi_R \bigg(8 a^5 \alpha R^8 + (1 + b R^4)^4 + 4 a^3 R^4 (1 + b R^4) \big(R^2 + \nonumber\\&& 12 \alpha (1 + b R^4)\big)+a^4 R^6 \big(R^2 + 32 \alpha (1 + b R^4)\big)
   +\beta \big(1 + b R^2 (4 \alpha + R^2) (2 + b R^4)\big)+2 a (1 + b R^4)\times\nonumber\\
   && \Big[R^2 \big(\beta +2 (1 + b R^4)^2\big)+4 \alpha \big(\beta + (1 + b R^4)^3\big)\Big] +a^2 R^2 \Big[R^2 \big(\beta + 6 (1 + b R^4)^2\big) +4 \alpha \big(\beta + 8 (1 + b R^4)^3\big)\Big]\bigg)\Bigg] \label{b2}
\end{eqnarray}
\end{widetext}
Now by solving Eqn. (\ref{ex2}), we obtain the expression for $C$ in terms of $B$ as,
\begin{eqnarray}
C  &=& e^{-\frac{BR^{2}}{2}} {\cal {F}}(R), \label{ex2a}
\end{eqnarray}
Here we have used the approximated mass and radius of the compact star candidate EXO 1785-248 to determine the values of $b, C, B$ and $\gamma$. To design a realistic stellar model, we fix the values of $\beta$ and $a$ as $\beta=0.25$ and $a=0.031$ with $K=0.29$ so as to obtain physical validity. Using these determined constants, we will now discuss the physical characteristics of our obtained solution.

\section{Physical analysis of the present charged model}\label{Sec6}
The model must satisfy the following regularity and reality requirements within the stellar configuration in order to be a viable, regular, and stable charged anisotropic compact stellar model with a certain radius. The behavior of anisotropic solutions of our compact stellar model is investigated through graphical analysis of metric potentials, matter variables, energy conditions, causality conditions, Herrera's cracking approach, adiabatic index, compactness factor, and redshift parameters.

\subsection{Regularity of metric co-efficient}
\begin{figure}[H]
    \centering
        \includegraphics[scale=.465]{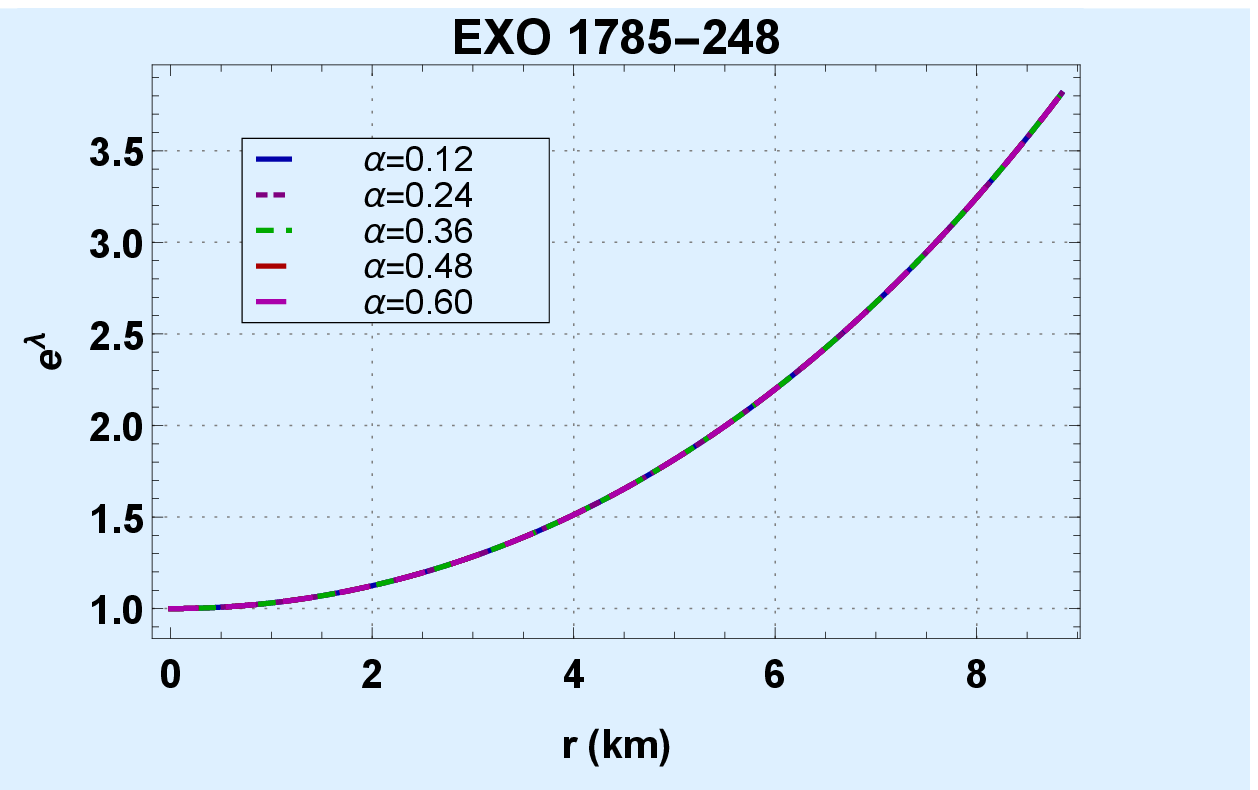}
        \includegraphics[scale=.465]{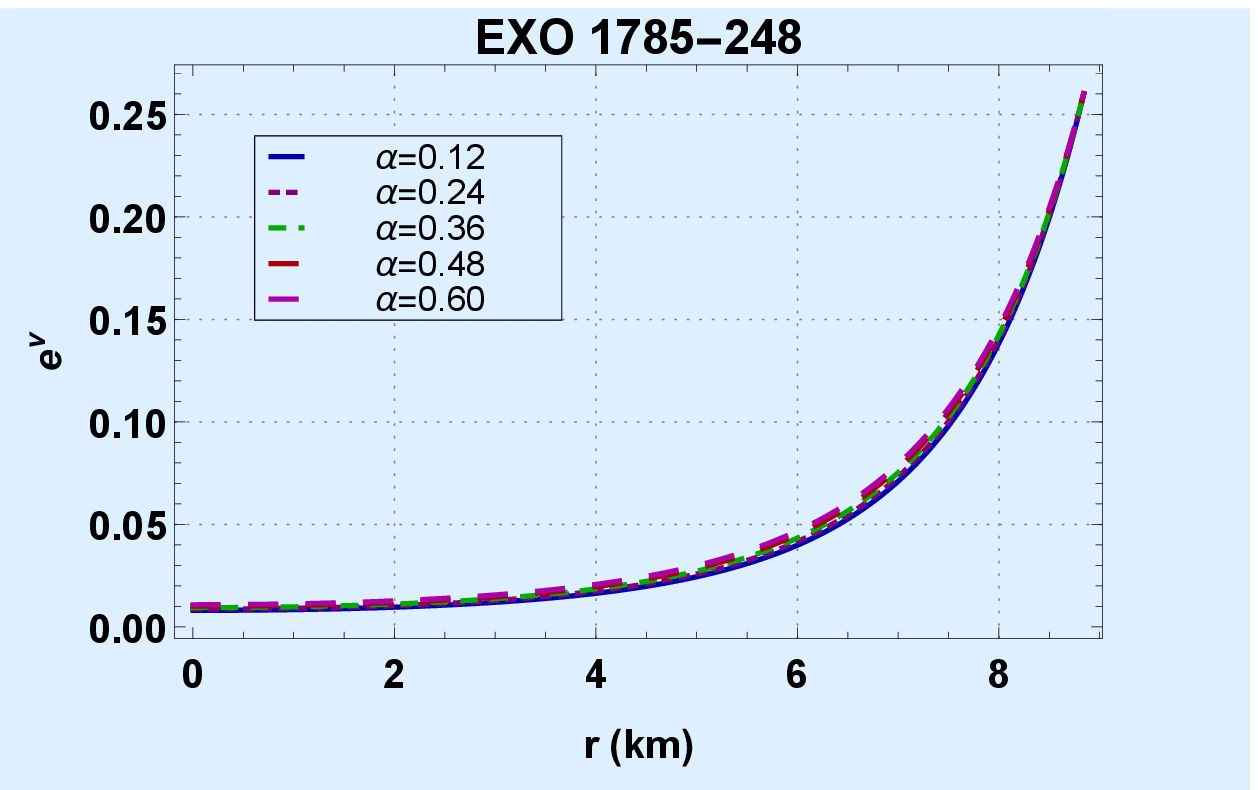}
        \caption{Variation of metric functions $e^{\lambda}$ and $e^{\nu}$ with respect to $r$ for EXO 1785-248. }
    \label{metric}
\end{figure}

In the survey of relativistic compact stars, the presence of singularities is recognized as a crucial problem. With this in mind, we are investigating the nature of the metric constituents in the core of these compact relativistic stars. For the solutions to be physically realizable \cite{Lake:2003}, the metric constituents inside the compact stars must be regular and positive. From Eq. (\ref{tk}), we can see that
\begin{eqnarray}
\frac{d}{dr}\left[e^{\nu}\right]=2BC^2 r e^{B r^2},~~\frac{d^2}{dr^2}\left[e^{\nu}\right]=\left(1+2 r^2\right)e^{B r^2}.\label{tk'}
\end{eqnarray}
This gives $\frac{d}{dr}\left[e^{\nu}\right]_{r=0}=0$ and $\frac{d^2}{dr^2}\left[e^{\nu}\right]_{r=0}>0$ at the center. We can also notice in Eq. (\ref{tk}) that the other metric element ensures the form $\left[e^{\lambda}\right]_{r\ne0}=1+O(r^2)$ near the center. Therefore, the metric components considered are non-singular at $r = 0$ and monotonically increasing. They thus satisfy all the conditions required to be physically acceptable which can be verified from Fig. (\ref{metric}).
\subsection{Regularity of pressure and density}
To reinforce the physical validity of our solution, the central values of the other physical quantities, namely matter density, and pressure must also be finite and non-singular to show their finiteness at the center. In this context, we found the central values of the physical quantities at the center $r = 0$. Firstly, the central density is obtained as,
 \begin{eqnarray}
 \rho_c=\rho(r=0)&=& \frac{3 (1 + 8 a \alpha) (a + B) + 4 \gamma \pi}{4 (1 + \beta) \pi}, \label{rc1}
 \end{eqnarray}
and secondly, the expression for central pressure is obtained as,
\begin{eqnarray}
p_c &=& p_r(r=0)= \frac{3 (1 + 8 a \alpha) (a + B)\beta - 4 \gamma \pi}{4 (1 + \beta) \pi}.\label{rc2}
 \end{eqnarray}
The radial and tangential pressures are equal and always positive at the center, where $r = 0$. A graphical representation of the requirements of the star enables for testing if they are always satisfied. 
\begin{figure}[H]
   \centering
        \includegraphics[scale=.465]{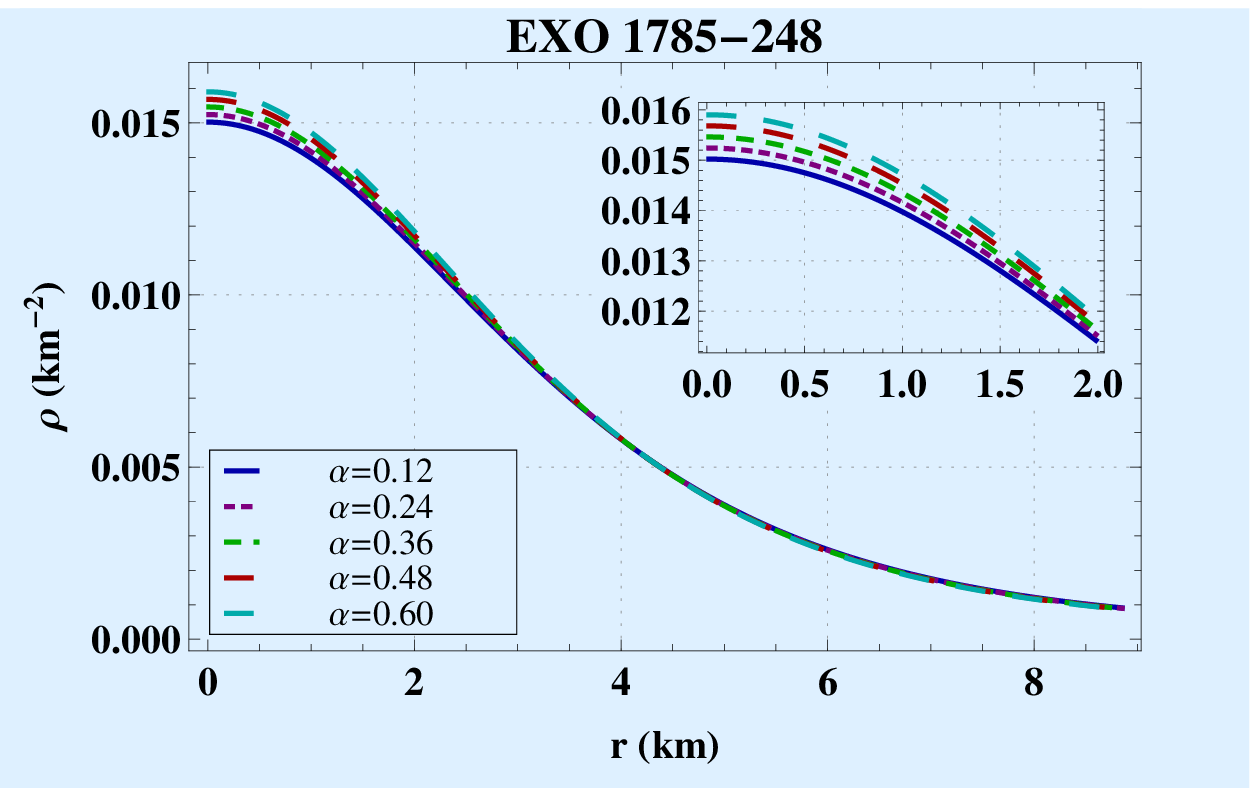}
        \includegraphics[scale=.465]{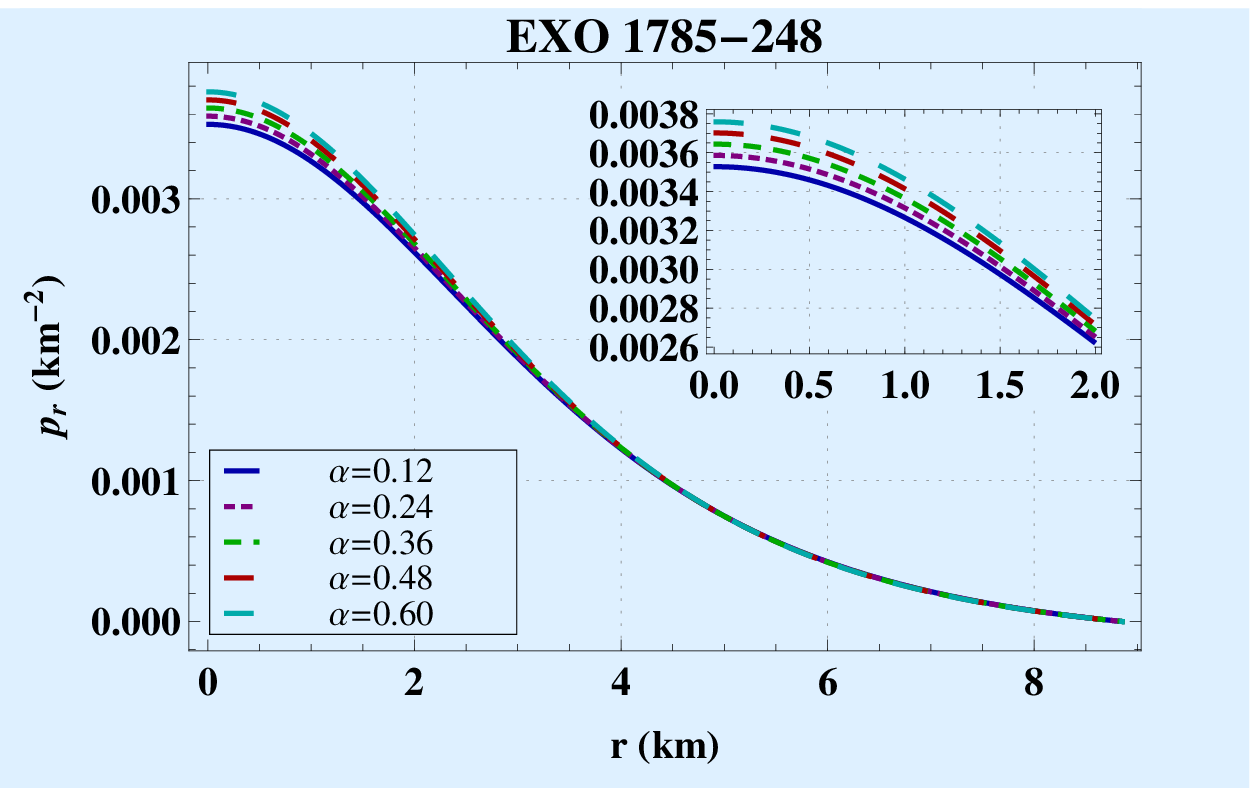}
        \includegraphics[scale=.465]{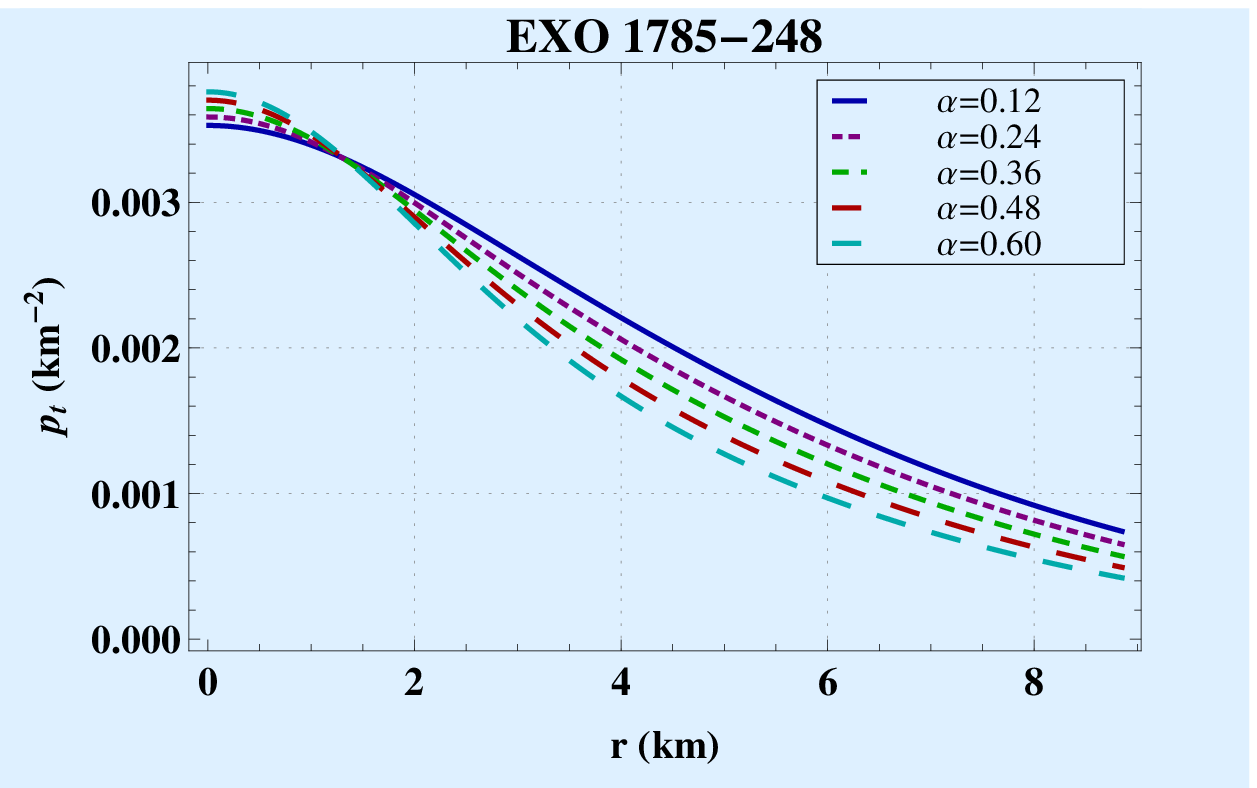}
       \caption{(Left) Matter density ($\rho$), (Middle) Radial pressure ($p_r$) and (Right) Tangential pressure ($p_t$) are plotted against the radial distance $r$.} \label{rho}
\end{figure}
\subsection{Evolution of fluid parameters}
\begin{figure}[H]
    \centering
        \includegraphics[scale=.465]{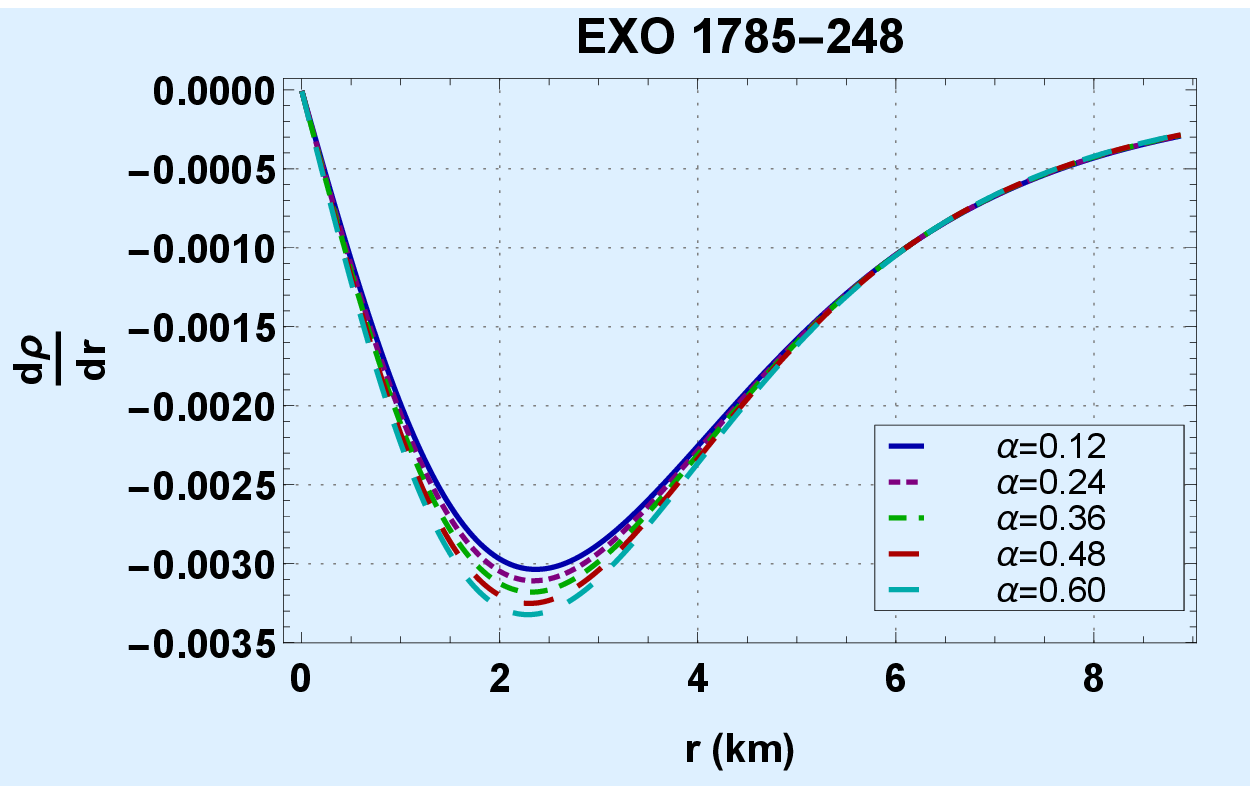}
        \includegraphics[scale=.465]{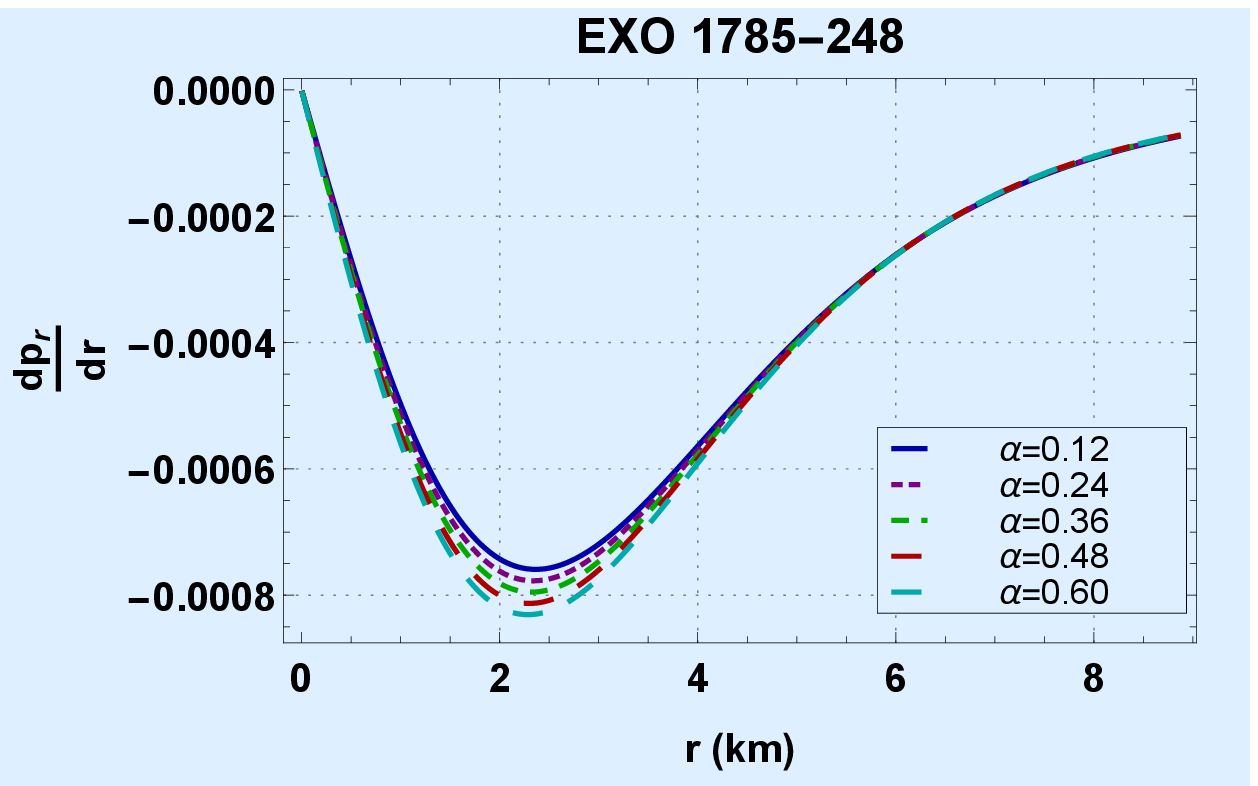}
        \includegraphics[scale=.465]{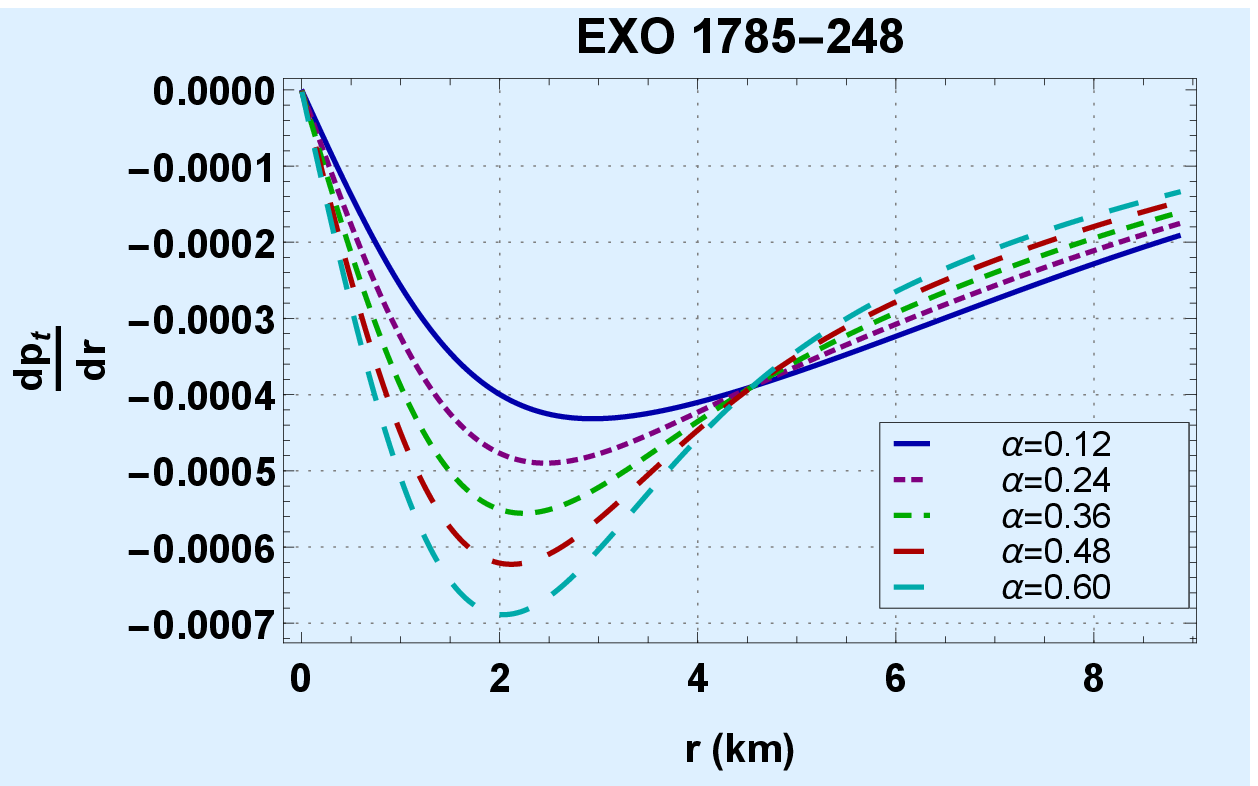}
       \caption{Density and pressure gradients: (Left) $\frac{d\rho}{dr}$, (Middle) $\frac{dp_r}{dr}$ and (Right) $\frac{dp_t}{dr}$ are shown against the radial distance $r$.}
    \label{dsvt25}
\end{figure}

Here we explore the viable properties of charged anisotropic compact stellar configuration and describe its evolution graphically. Furthermore, the comparison of our theoretical findings with empirical data could provide strong evidence in favor of the considered 5-D EMGB model. We start by analyzing some realistic characteristics of the stellar configurations such as the matter density, radial and lateral pressures along with their derivatives, the anisotropy factor, and the trace profile through graphs. We acknowledge that due to the extremely dense stellar profile, the matter contents should be higher in the stellar interior. In this respect, we have shown in Fig. \ref{rho} that the behavior of the physical parameters, namely $\rho$, $p_r$ and $p_t$ is positive and decreases monotonically towards the stellar surface, which shows a very compact profile of the considered stellar configuration. Then, we analyzed the density and pressure gradients viz., ${d\rho}/{dr}$, ${dp_r}/{dr}$, ${dp_t}/{dr}$ and demonstrated that their behaviors are negative and regular, which ensures the presence of an intensive stellar object in the 5-D EMGB model, as illustrated in Fig. \ref{dsvt25}. It's worth noting that the matter variables have high values compared to the GTR.

\begin{figure}[H]
    \centering
        \includegraphics[scale=.465]{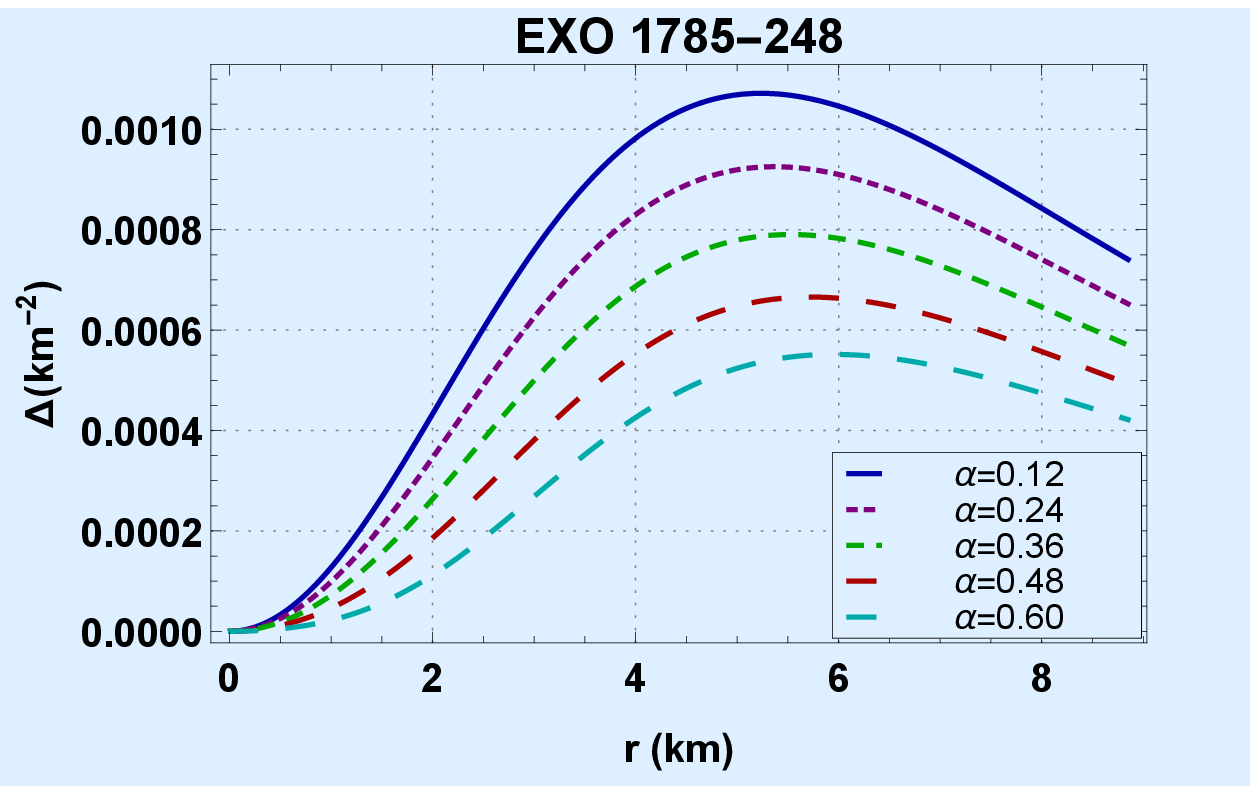}
         \includegraphics[scale=.465]{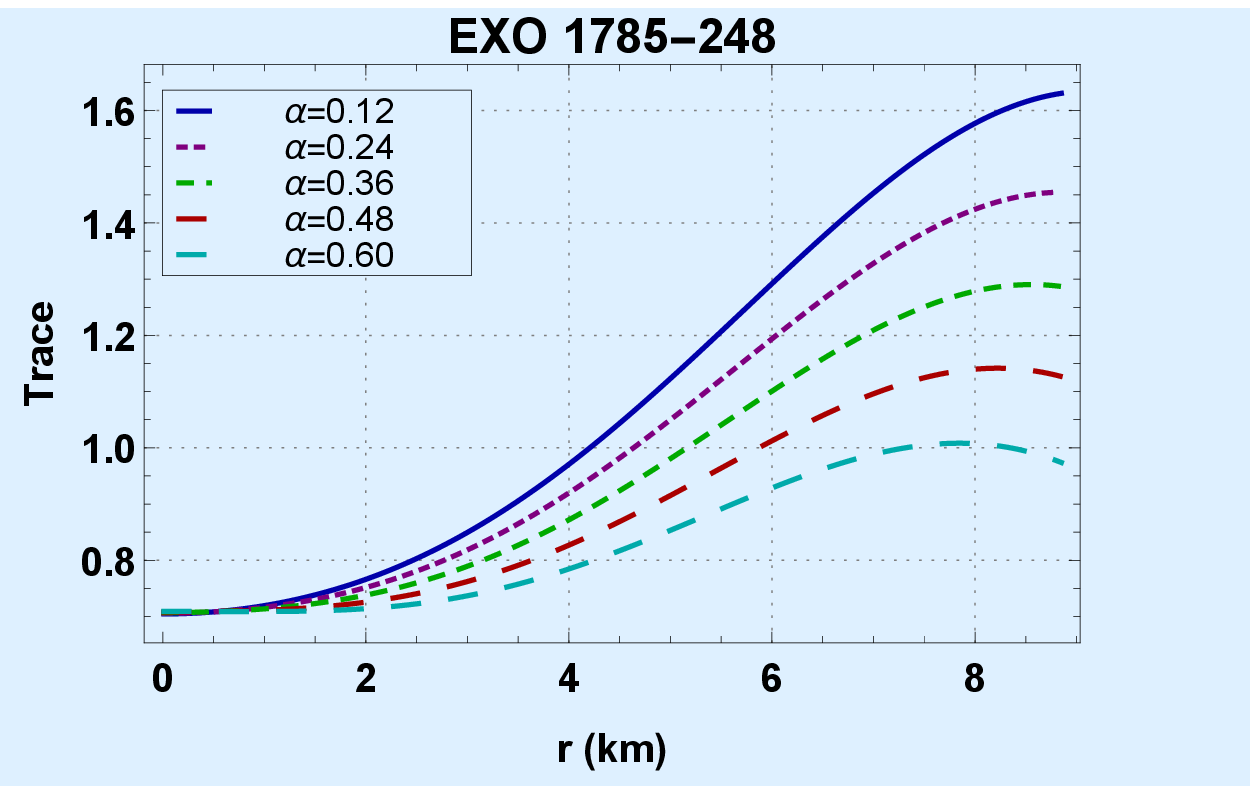}
       \caption{(Left) Anisotropic factor ($\Delta$) and (Right) Trace profile ($\frac{p_r+2p_t}{\rho}$) are plotted against the radial distance $r$.}
\label{delta}
\end{figure}

An interesting key factor when analyzing the matter configuration is the pressure anisotropy, whose direction relies on the pressure components. When $\Delta = p_t - p_r < 0$, the anisotropy is negative, indicating that the pressure is oriented internally, while when $\Delta = p_t - p_r > 0$, the anisotropy is positive, indicating that the pressure is oriented externally. The graphical visualization of the anisotropy factor $\Delta$ is yielded in Fig. \ref{delta} (left panel), which shows that the current stellar model has a positive anisotropy factor $\Delta$, where $p_{t}>p_{r}$ then $\Delta>0$. Therefore, the stellar structure is disclosed to a repulsive force that offsets the gravitational gradient, this phenomenon enables the production of a more compact stellar configuration. The radial and transverse pressures coincide at the center, i.e., $\Delta$ vanishes at $r=0$. Additionally, the radial and transverse pressure values diverge with increasing radius, which causes the anisotropy to keep moving in the direction of the surface of the spherical stellar configuration. Besides, Fig. \ref{delta} (right panel) demonstrates that the trace profile i.e., ($\frac{p_r+2p_t}{\rho}$) is positive with increasing behavior throughout the interior of the object. This result advocates the hefty profile of stellar matter variables, thereby representing compact environments of the considered stellar configuration.

\subsection{Electric field}
\begin{figure}[H]
    \centering
        \includegraphics[scale=.465]{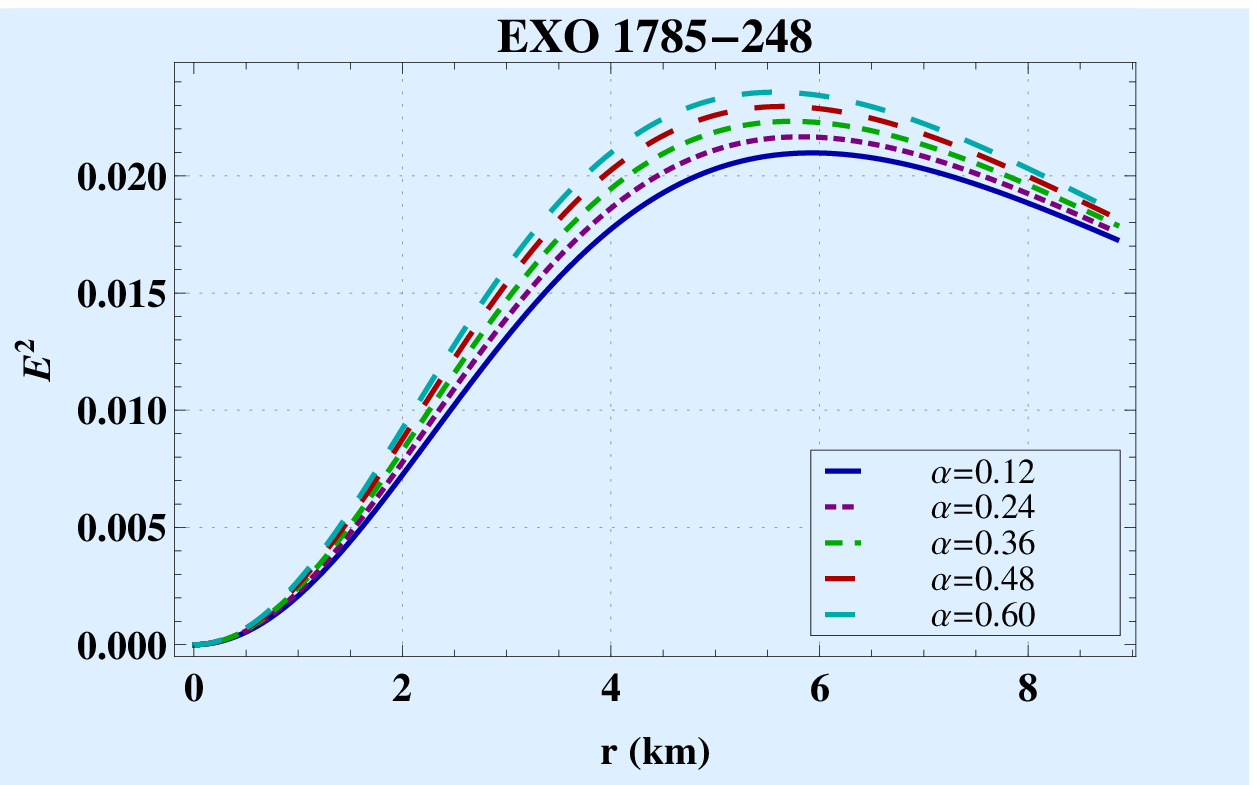}
        \includegraphics[scale=.465]{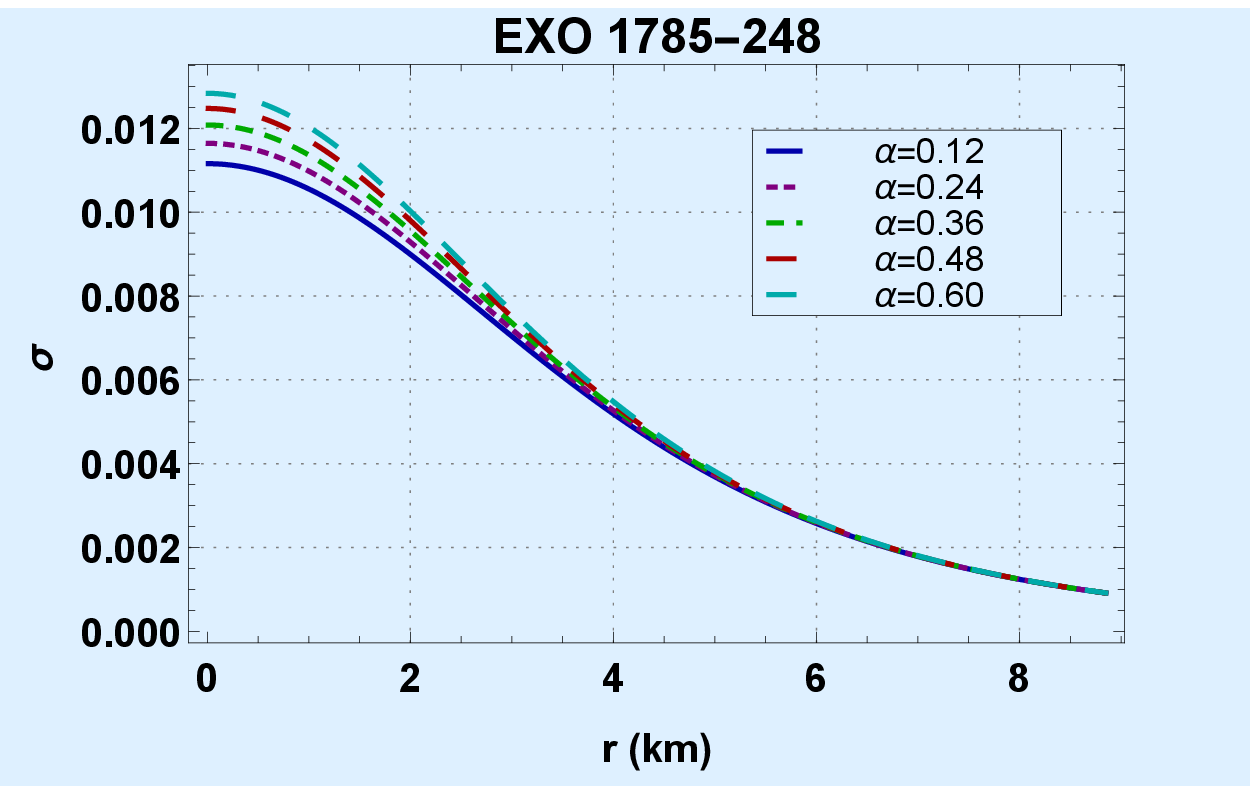}
        \caption{(Left) Total charge ($E^2$) and (Right) Charge density ($\sigma$) are shown against the radial distance $r$. } \label{ener}
\end{figure}

It is generally recognized that the regularity of the electric charge must be effectively zero at the center and positive for $r > 0$, which aids in avoiding gravitational collapse. To confirm this regularity, we have presented in Fig. \ref{ener} (left panel), the trend in the electric field. In this context, we notice that the electric field vanishes at the stellar core and is positive throughout the stellar configuration, but it takes a maximum value within the stellar configuration at $r\approx 5.5~km$ instead of the stellar surface. It is well-known that the stellar core can become unstable under the effect of strong electric fields. On the other hand, quasi-static equilibrium states can be generated by the presence of charges up to $10^{20}$ coulombs. Due to pair production within the stellar configuration caused by very high charge densities and extremely strong electric fields, the core of the stellar configuration becomes unstable. Next, Fig. \ref{ener} (right panel) illustrates how the charge density behaved for the hypothetical stellar configuration. The charge density is consequently seen to be non-singular at the stellar configuration's center, decreases monotonically to the stellar configuration's surface layer, and attains a non-zero value at the stellar frontier.

\subsection{Energy conditions}
\begin{figure}[H]
    \centering
        \includegraphics[scale=.465]{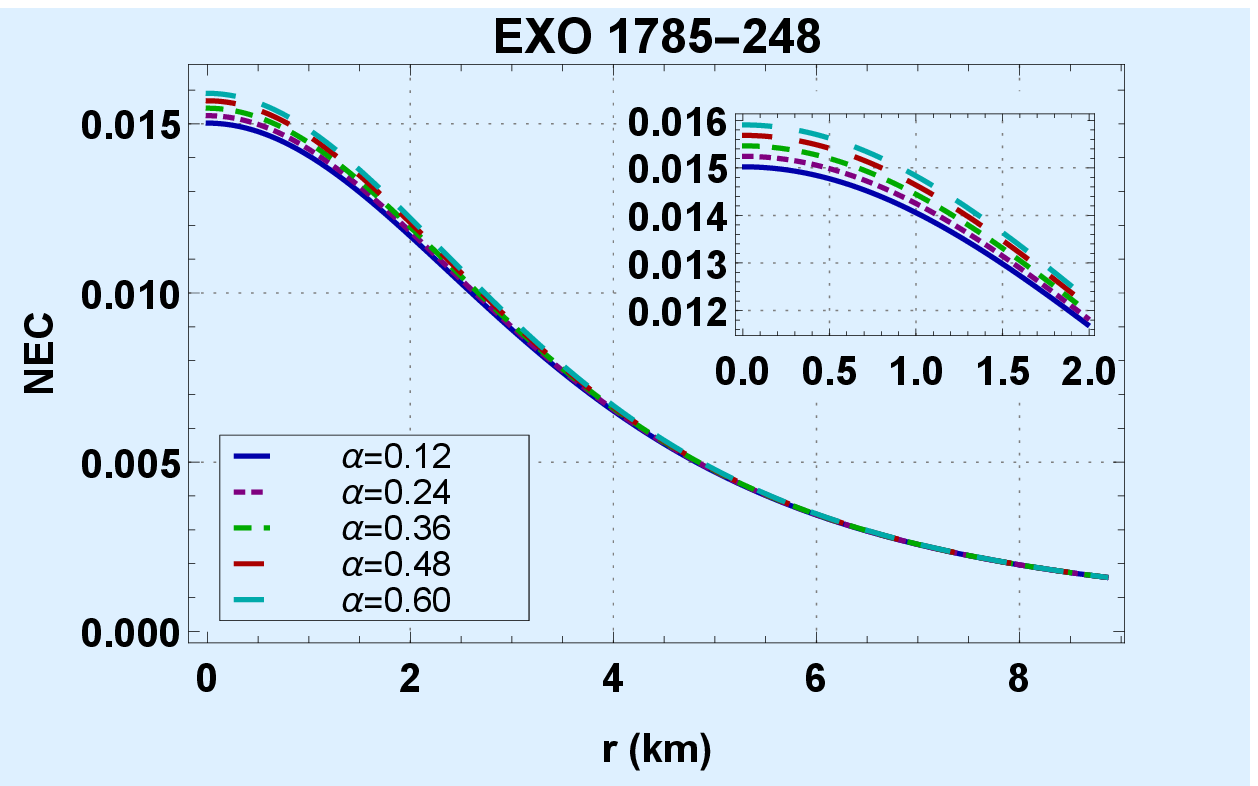}
        \includegraphics[scale=.465]{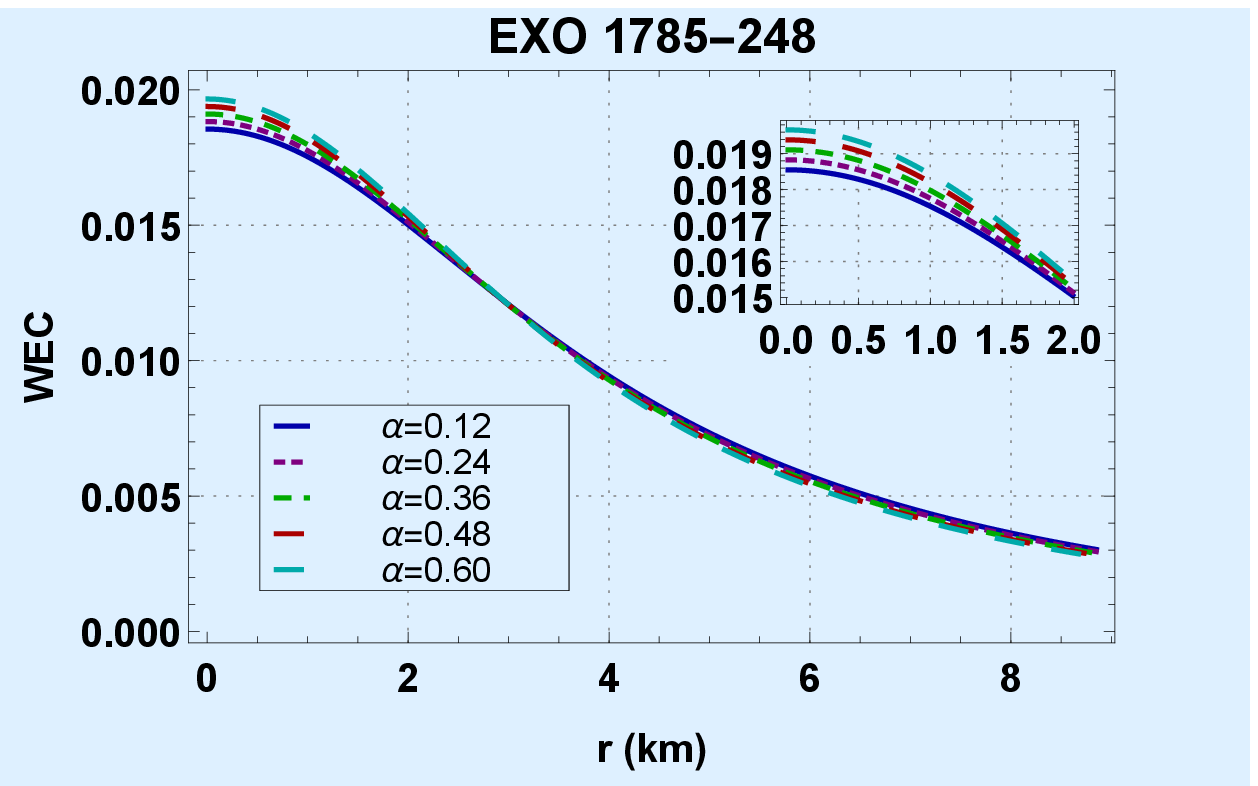}
        \includegraphics[scale=.465]{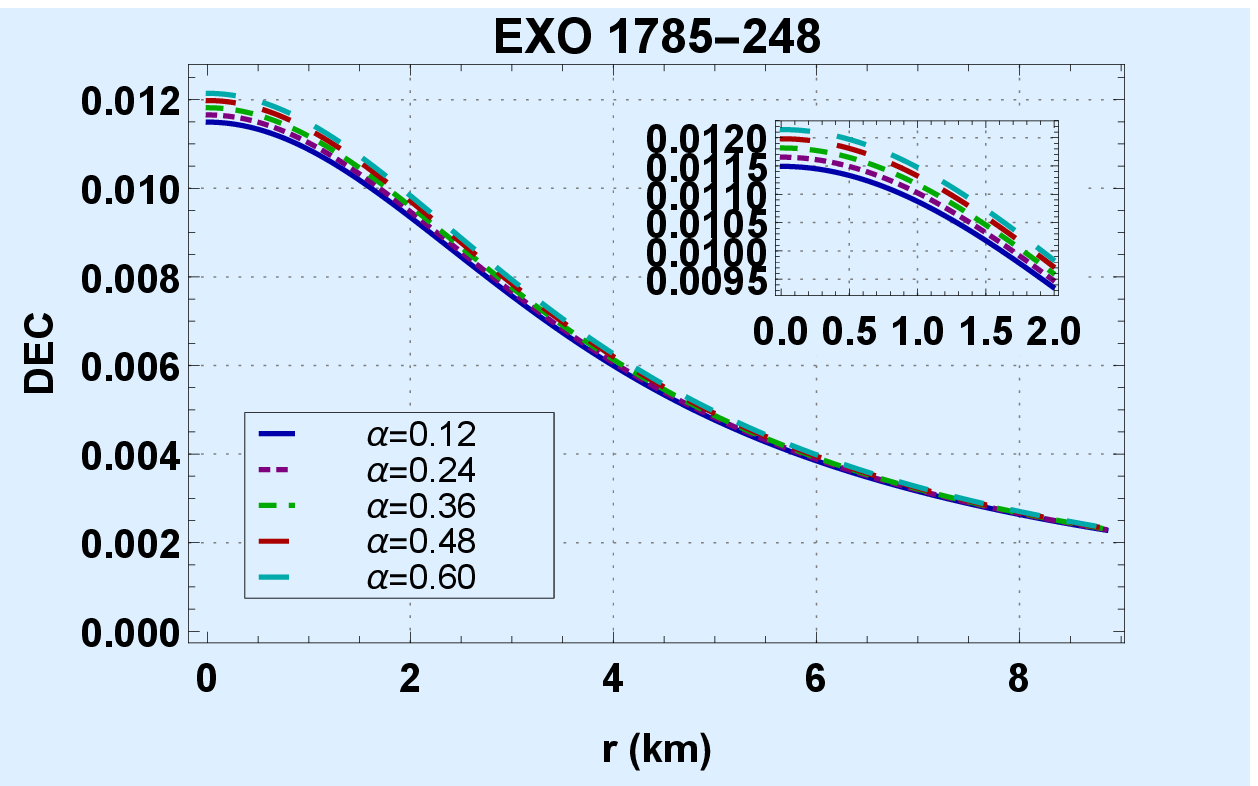}
        \includegraphics[scale=.465]{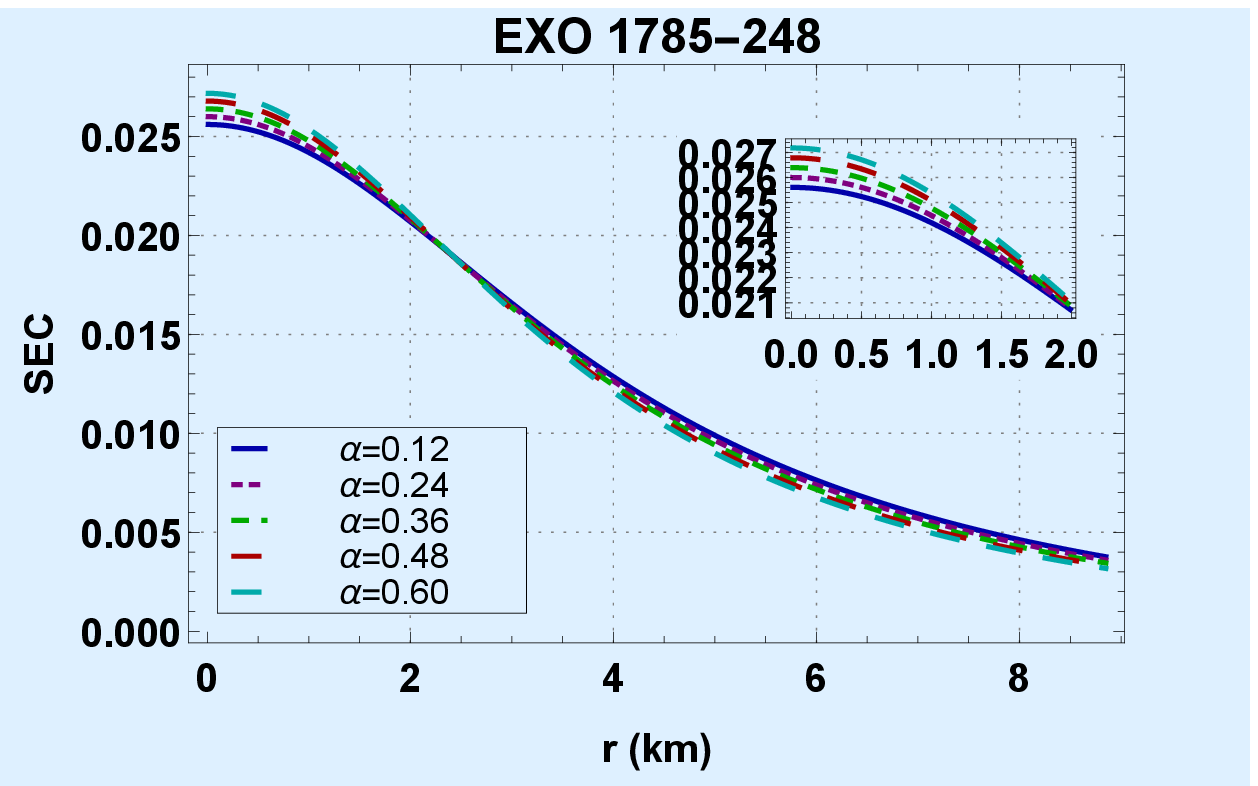}
       \caption{Variation of energy conditions with respect to the radial distance $r$.}
    \label{ec1}
\end{figure}

An ordinary matter can be distinguished from an exotic fluid on the basis of the energy conditions (ECs) on matter variables. These restrictions give detailed information about the physical viability of the solutions. Strong, weak, null, and dominant energy conditions (SEC, WEC, NEC, and DEC) for charged stellar configuration are given by the following inequalities,
\begin{eqnarray}
    SEC&:&\rho+p_r+2p_t+\frac{E^{2}}{4\pi}\geq0,\label{53}\\
    WEC&:& \rho+p_r\geq0,~~ \rho+p_t+\frac{E^{2}}{4\pi}\geq0\label{54}\\
    NEC&:&\rho+\frac{E^{2}}{8\pi}\geq0,\label{55}\\
    DEC&:&\rho-p_r+\frac{E^{2}}{4\pi}\geq0, ~~ \rho-p_t\geq0.\label{56}
    \end{eqnarray} 
Fig. \ref{ec1} illustrates the graphical behavior of these energy conditions which shows that our system is consistent with all these ECs. This guarantees that the charged matter that composes our stellar model is realistic and physically viable.

\subsection{Stability of the stellar model}\label{Sec7}
In this section, we will investigate the stability of our stellar model by using the two methods given by (i) Causality condition via Herrera's cracking concept and (ii) determining the relativistic adiabatic index.

\subsubsection{Causality condition via Herrera's cracking concept}

\begin{figure}[H]
    \centering
        \includegraphics[scale=.41]{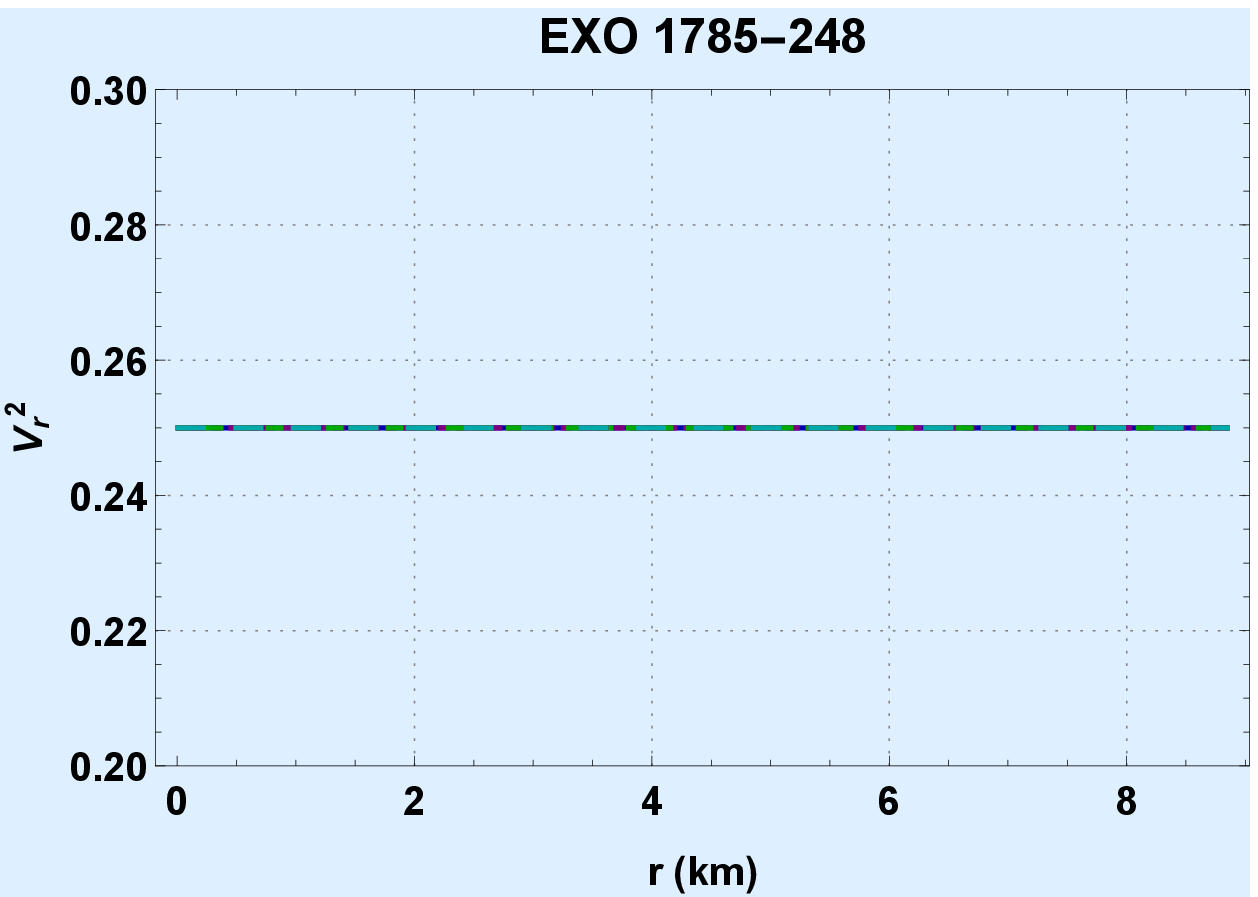}
        \includegraphics[scale=.465]{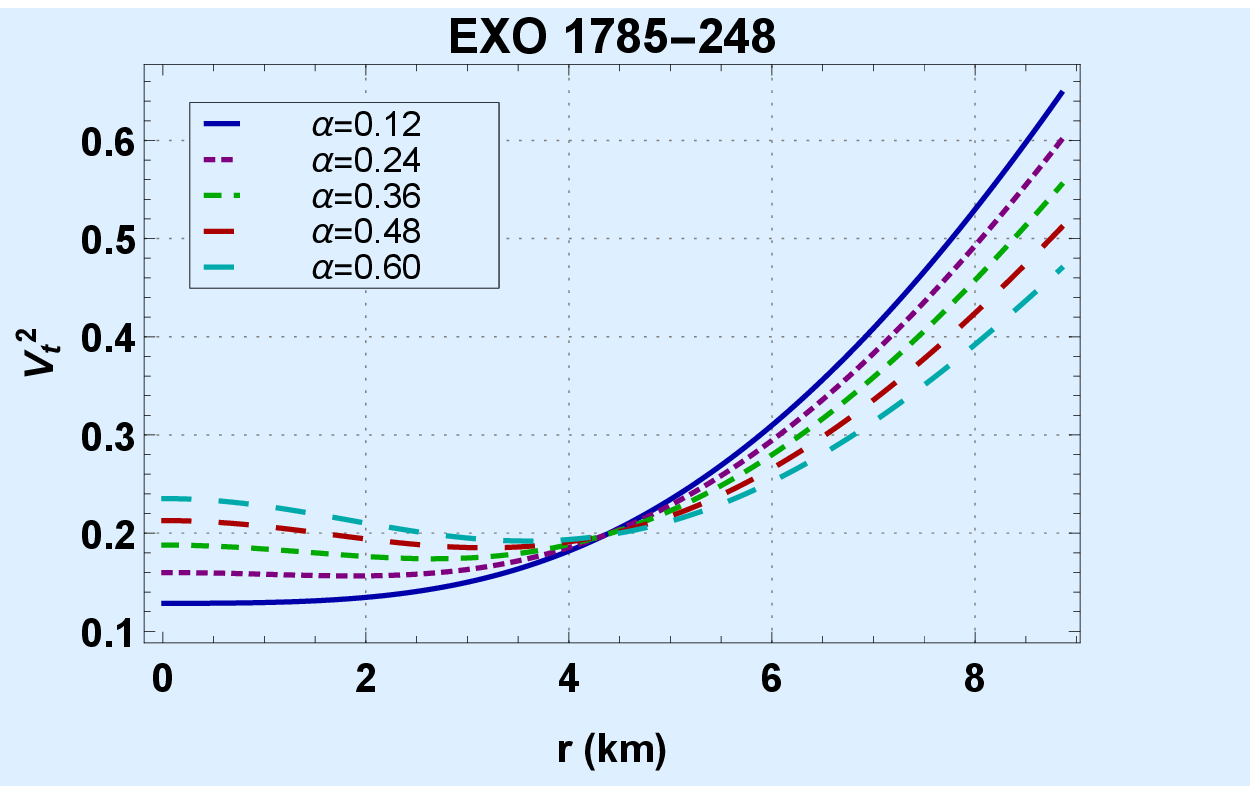}
        \includegraphics[scale=.465]{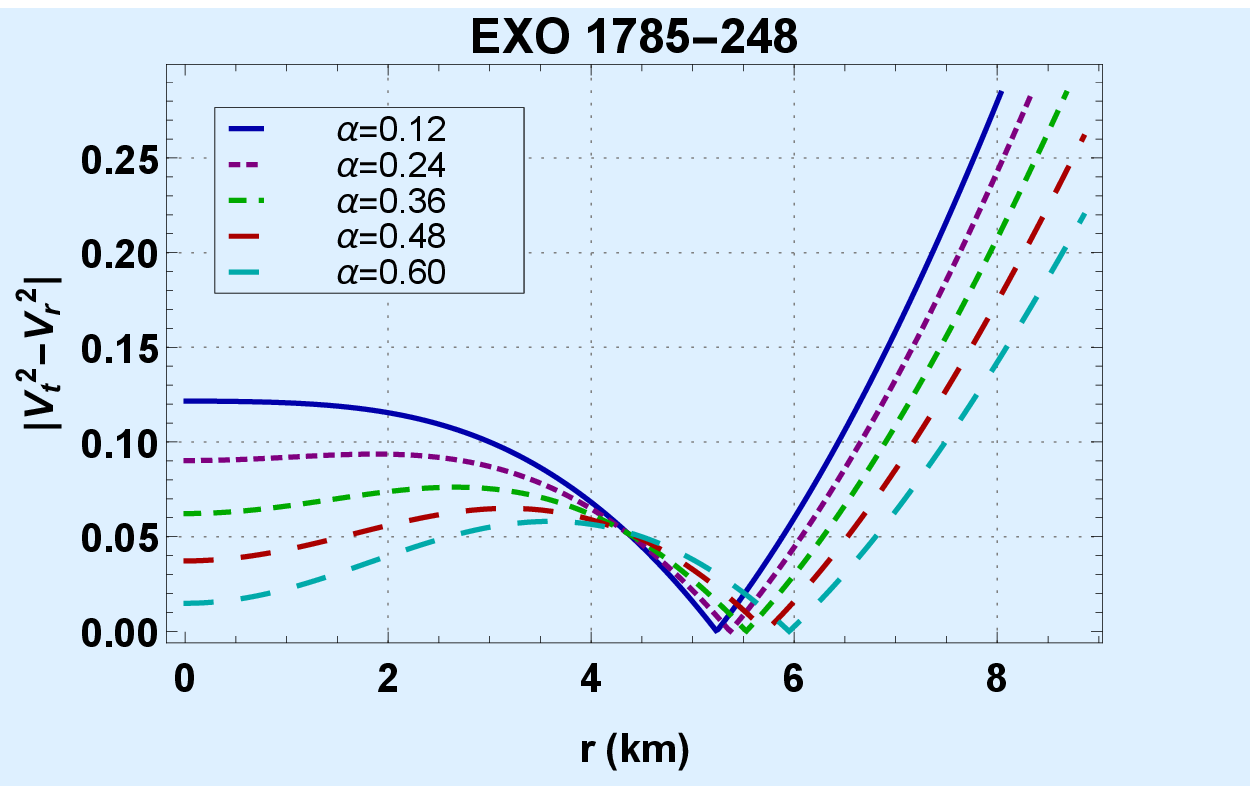}
       \caption{Components of sound velocity: (Left) $v_r^2$ is plotted against r, (Middle) $v_t^2$ is shown against the radial distance $r$ and (Right) the stability factor $|v_t^2 - v_r^2| $ is plotted against $r$. }
    \label{sv}
\end{figure}

A very important condition to be a physically acceptable stellar model is the causality condition, which states that the square of the radial sound velocity ($v_r^2 = \frac{dp_r}{d\rho}$) and that for tangential sound velocity ($v_t^2 = \frac{dp_t}{d\rho}$) for anisotropic compact object should be less than unity everywhere \cite{abr}, i.e.,
\begin{eqnarray}
      0< v_r^2 &=& \frac{dp_r}{d\rho} < 1 \nonumber\\
      0< v_t^2 &=& \frac{dp_t}{d\rho} < 1
\end{eqnarray}
In other words, these conditions state that the velocity components of sound must be less than the speed of light. Since we have assumed a linear equation of state for radial pressure $p_r$, then from
Eqn. (\ref{eos}), we obtain $v_r ^2 = \beta$. Therefore, the radial velocity component is constant throughout the stellar model and totally depends on parameter $\beta$, while tangential velocity component varies with parameter $\alpha$. The expression of $v_t ^2$ in EMGB gravity is highly non-linear and complicated that has to be analyzed numerically.
From Fig. \ref{sv}, it is very clear that the values of all velocity components lie within $[0,1]$. 

On the other hand, Herrera et al \cite{her1,her2,her3} proposed the cracking concept for self-gravitating compact stellar model by considering perfect fluid and anisotropic matter distributions. Following this concept, we have checked the cracking condition  $|v_t^2 - v_r^2| \leq 1$ and thus from Fig. (\ref{sv}) we find that the present compact stellar model is potentially stable by satisfying both the causality conditions and the Herrera's cracking concept.

\subsubsection{Relativistic adiabatic index}

The adiabatic index, which measures the stiffness of the matter, is crucial for understanding the dynamical stability of a relativistic compact object. Chandrasekhar first proposed the dynamical stability of stellar systems against infinitesimal radial adiabatic perturbation \cite{chandra}. In his pioneering work, he predicted that for stability of a stellar system the value of adiabatic index $\Gamma$ should be greater that $4/3$. Several researchers have tested the predictions of Chandrasekhar for both isotropic and anisotropic stellar objects \cite{ad1,ad2,ad3,ad4}. 
\begin{figure}[H]
    \centering
        \includegraphics[scale=.5]{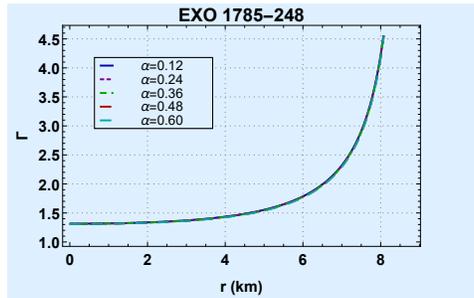}
       \caption{Behavior of the adiabatic index $\Gamma$  as a function of the radial distance $r$.}
    \label{gammar}
\end{figure}
The adiabatic index is defined as,

\begin{eqnarray} \label{ad}
\Gamma=\frac{\rho+p_r}{p_r}v_r^2. 
\end{eqnarray}

We have successfully plotted the behaviour of the adiabatic index $\Gamma$ in  Fig. (\ref{gammar}) which shows that $\Gamma$ is greater than $4/3$ for our proposed model. This confirms the stability of this stellar system.

\subsection{Hydrostatic Equilibrium via Modified TOV Equation}
The four different forces, namely gravitational force, hydrostatics force, electric force and anisotropic force are acting in our present model.
In this section, we will investigate the hydrostatic equilibrium of our proposed model via the modified
Tolman-Oppenheimer-Volkoff (TOV) equation under different forces. To analyze, we write the general form of the modified TOV equation as follows:
\begin{equation}\label{tov1}
-\frac{M_G(r)(\rho+p_r)}{r}e^{\nu-\lambda}-\frac{dp_r}{dr}+\frac{3}{r}(p_t-p_r)+\sigma Ee^{\lambda}=0,
\end{equation}
proposed by Tolman-Oppenheimer-Volkoff and named as {\em TOV} equation.\\
Where $M_G(r)$ is the gravitational mass within the radius $r$ derived from the Tolman-Whittaker
formula and the Einstein's field equations and is defined by
\begin{equation}\label{tov2}
M_G(r)=\frac{1}{2}re^{\lambda-\nu}\nu'.
\end{equation}

Using the expression for $M_G(r)$ in equation \eqref{tov1}, we get
\begin{equation}
-\frac{\nu'}{2}(\rho+p_r)-\frac{dp_r}{dr}+\frac{3}{r}(p_t-p_r)+\sigma Ee^{\lambda}=0.
\end{equation}

\begin{figure}[htbp]
    \centering
        \includegraphics[scale=.465]{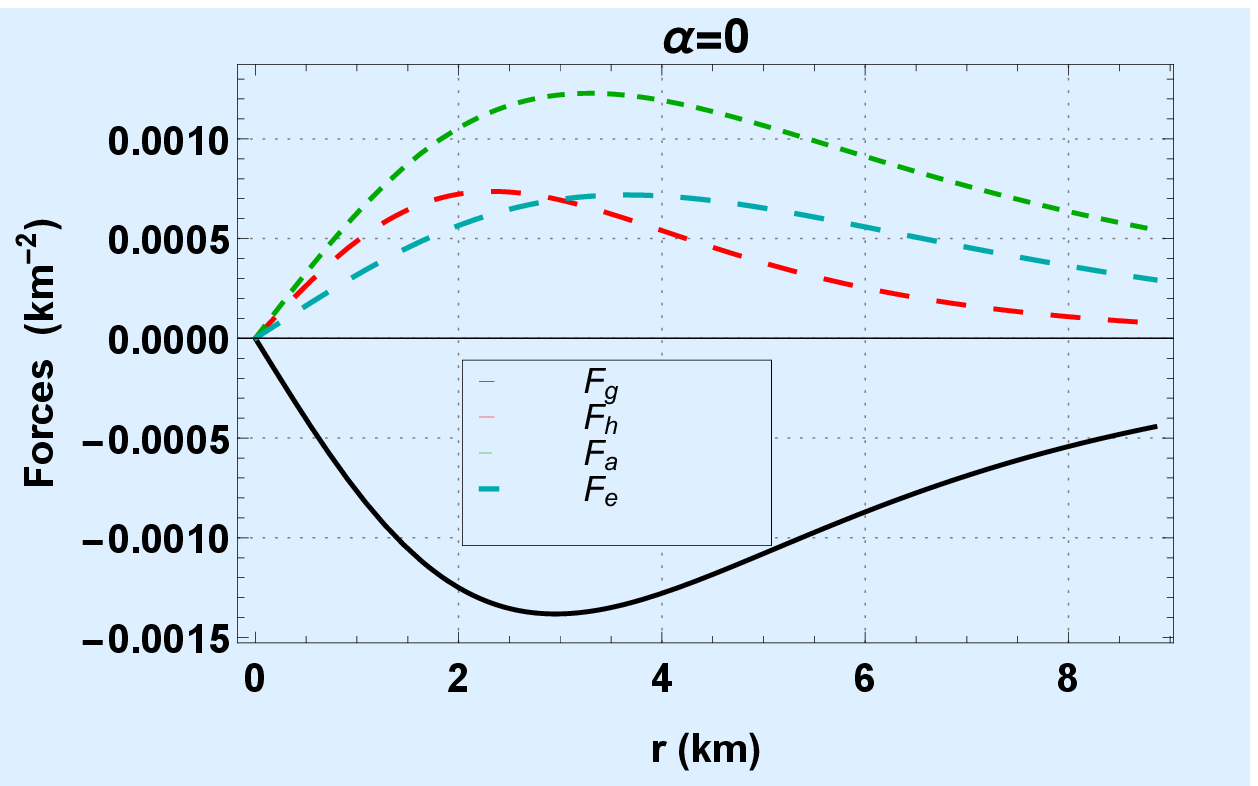}
        \includegraphics[scale=.465]{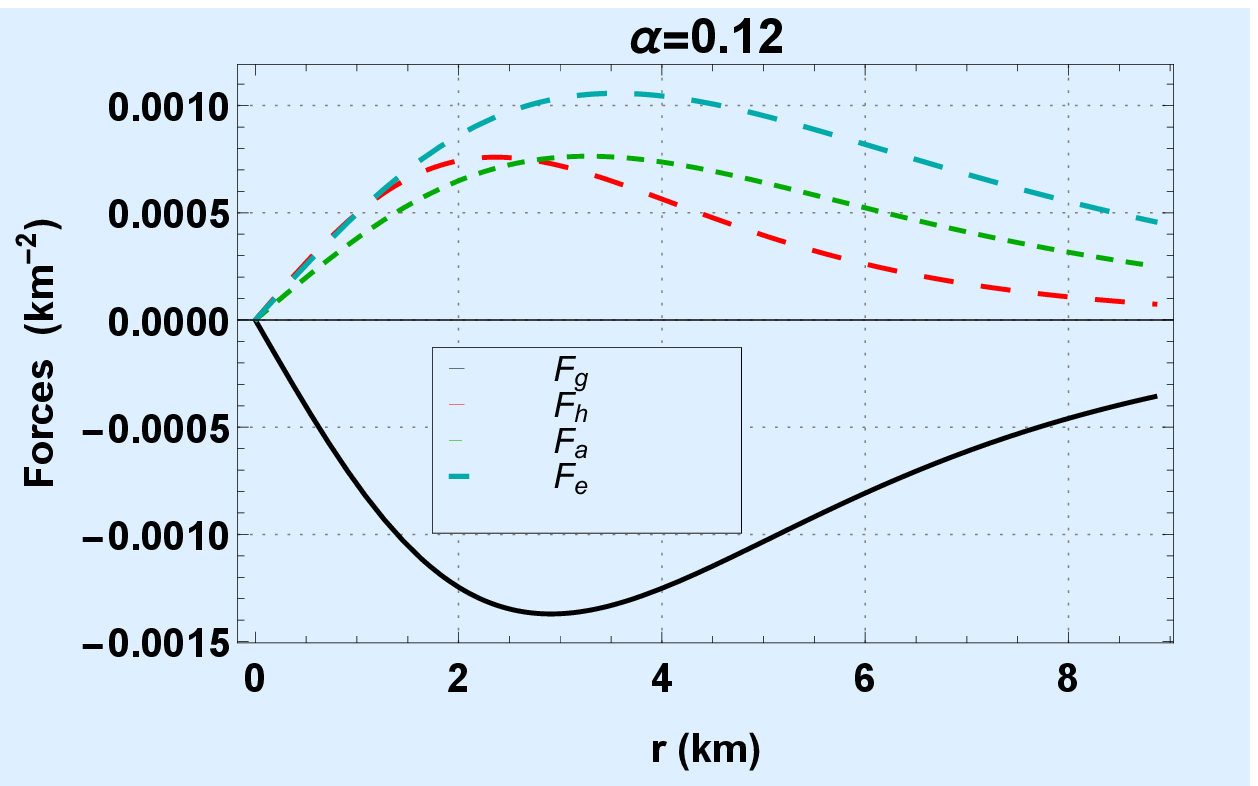}
        \includegraphics[scale=.465]{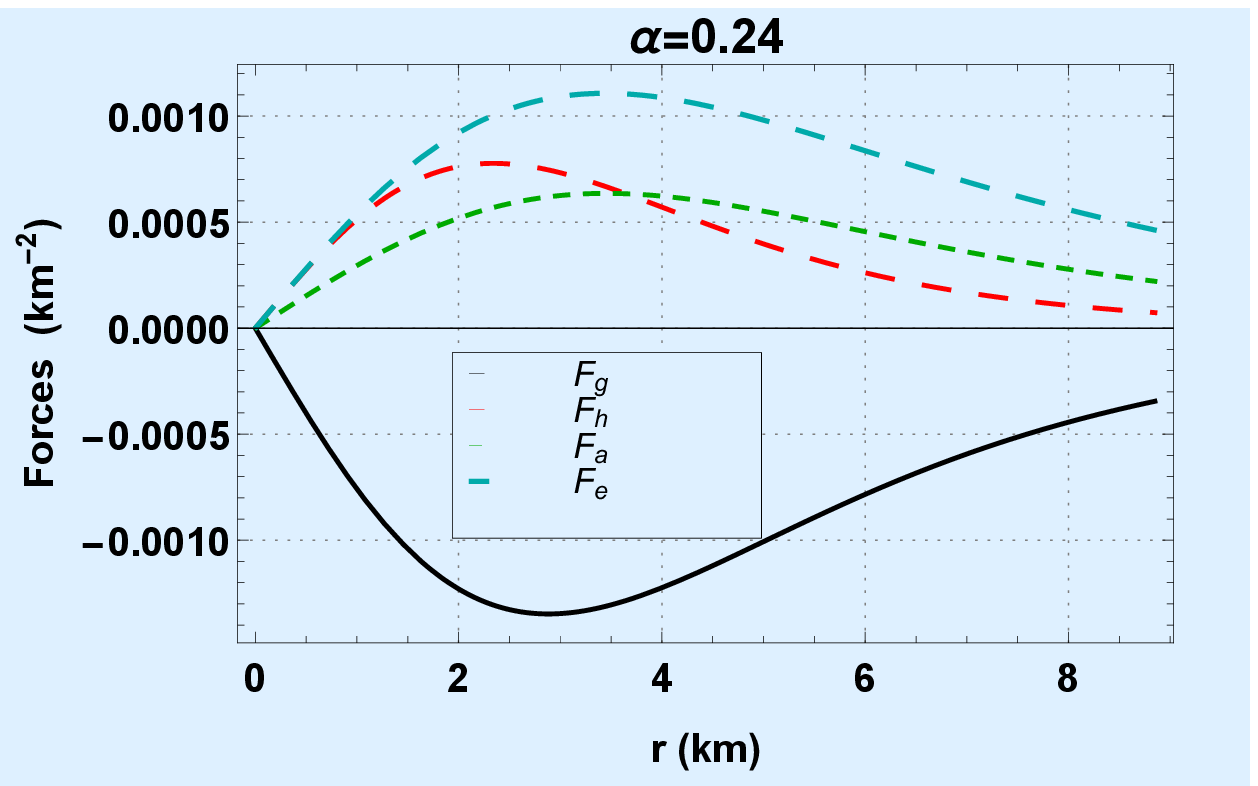}
        \includegraphics[scale=.465]{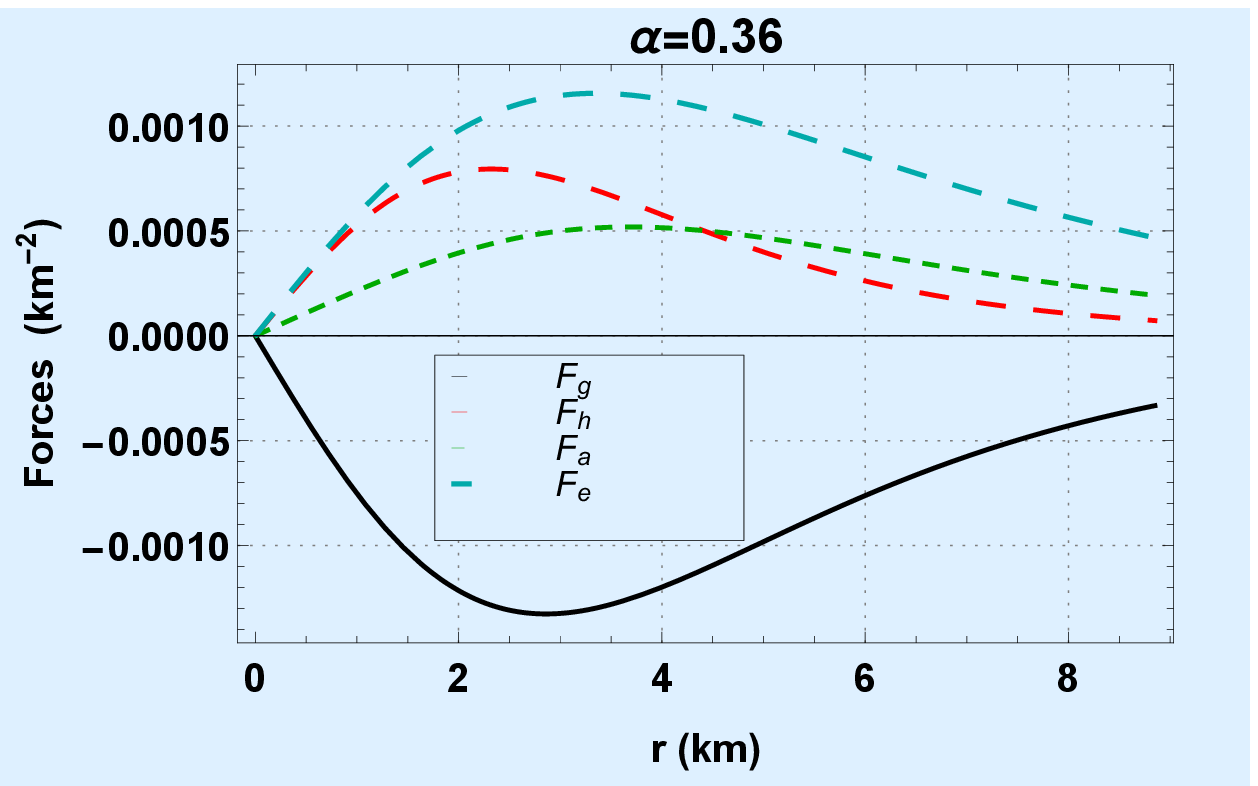}
        \includegraphics[scale=.465]{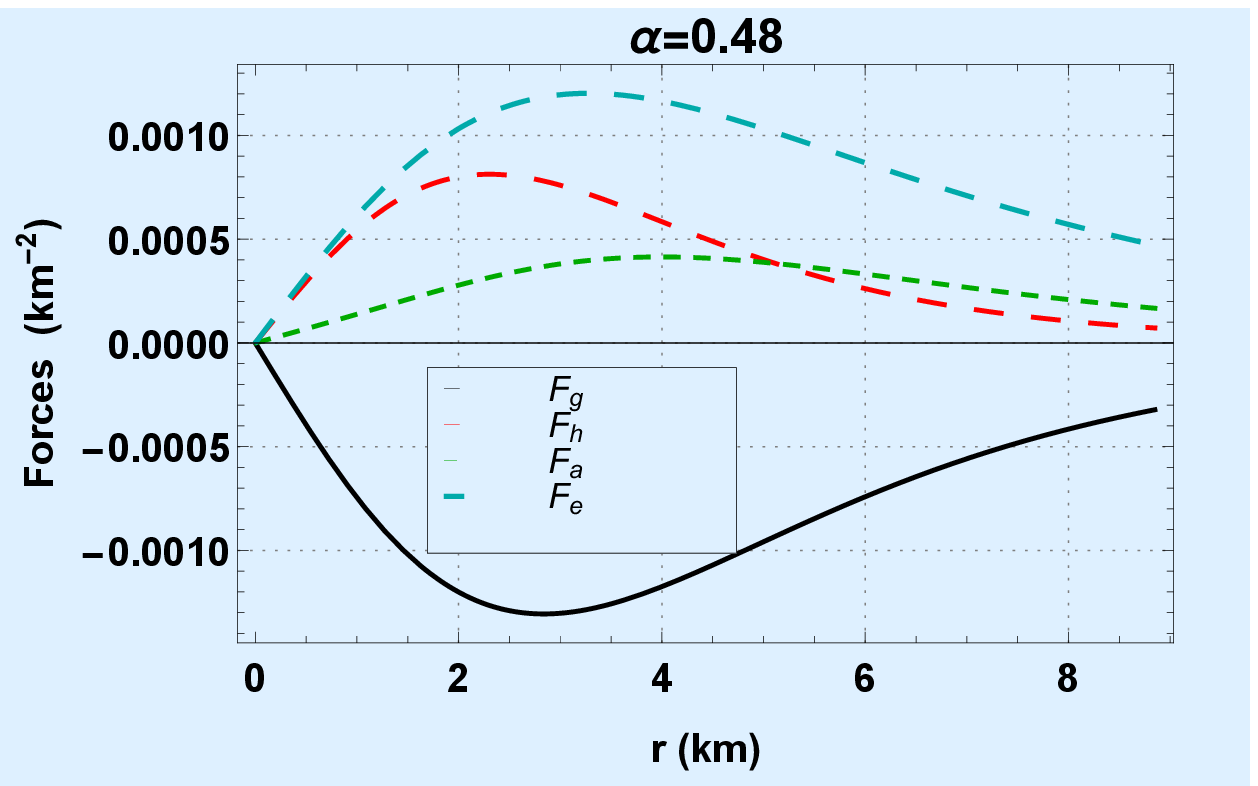}
        \includegraphics[scale=.465]{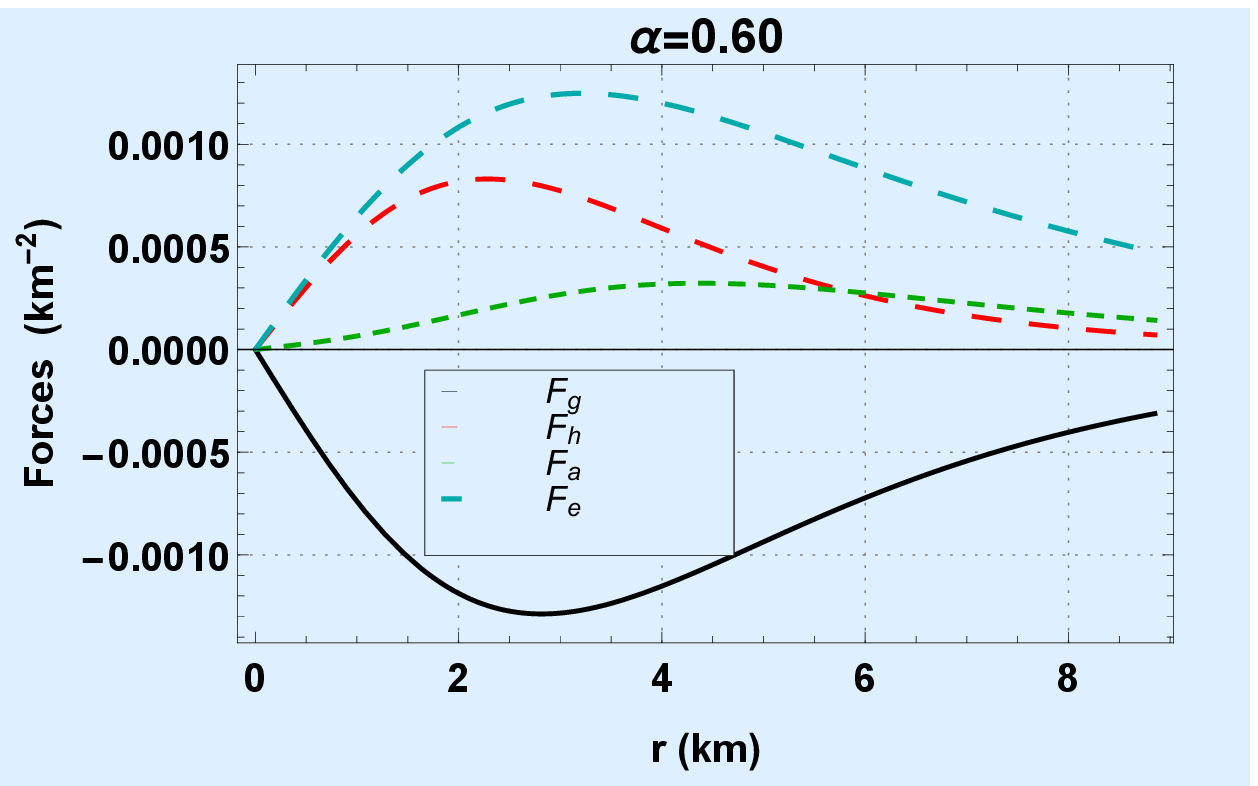}
       \caption{Variations of different forces, such as the gravitational force $F_g$, hydrostatic-gradient force $F_h$, anisotropic force $F_a$ and electric force $F_e$ respectively versus radial coordinate $r$ for different values of $\alpha$.}
    \label{tov}
\end{figure}

The above expression may also be written as
\begin{equation}
F_g + F_h + F_a + F_e=0,
\end{equation}
where $F_g, F_h$, $F_a$, and $F_e$ represent the gravitational, hydrostatic-gradient, anisotropic and electric force respectively given by the following expression.
\begin{eqnarray}
F_g&=&-\frac{\nu'}{2}(\rho+p_r) \\
F_h&=&-\frac{dp_r}{dr}\\
F_a&=&\frac{3}{r}(p_t-p_r)\\
F_e&=&\sigma Ee^{\lambda}.
\end{eqnarray}

\subsection{Some physical implications of the solutions for a toy stellar model}
we can readily measure the mass function inside the radius $R$ by computing the integral connected directly to the density level using the following definition,
\begin{eqnarray}
 m(R)=4\pi\int^R_0{\bigg(\rho+\frac{q^2}{8\pi\,r^4}\bigg)r^2\,dr}+\frac{Q^2}{2R},~~~\label{52}
 \end{eqnarray}
 conversely employing the metric function suggested by 
\begin{eqnarray}\label{53a} 
    m(R)=\frac{R}{2}\Big(1-e^{-2\lambda(R)}+\frac{Q^2}{R^2}\Big).
\end{eqnarray}
On the other hand, using the formula suggested by B\"ohmer and Harko \cite{Bohmer06}, it is easy to determine the lower and upper limits of the mass-radius ratio, 
\begin{eqnarray} \label{eq53b}
\frac{Q^{2}\left(18\,R^{2}+Q^{2}\right)}{2\,R^{2}\left(12\,R^{2}+Q^{2}\right)} \leq \frac{{M}}{R}\leq \frac{2}{9}+ \frac{3Q^{2}+2R\sqrt{R^{2}+3Q^{2}}}{9R^{2}},~~
    \end{eqnarray}
In addition, one may get the effective mass for the distribution of charged matter by,
\begin{eqnarray} \label{eq72}
&&\hspace{-0.6cm} M_{\textrm{eff}} =4\pi\int_{0}^{R}\left({\rho}+\frac{q^{2}}{8\pi\,r^4}\right)r^{2}dr =\frac{R}{2}\left[1-e^{-2\lambda(R)}\right].
\end{eqnarray}
We also signify compactification for the effective mass-to-radius ratio, $u$, which is given as follows,
\begin{eqnarray}
    u=\frac{M_{\textrm{eff}}}{R}.\label{53c}
\end{eqnarray}
This serves as the basis for evaluating the surface redshift function $z_s$, which is stated as,
\begin{eqnarray}
    z_s=\left(1-2u\right)^{-\frac{1}{2}}-1.\label{54a}
\end{eqnarray}

\begin{figure}[H]
    \centering
        \includegraphics[scale=.37]{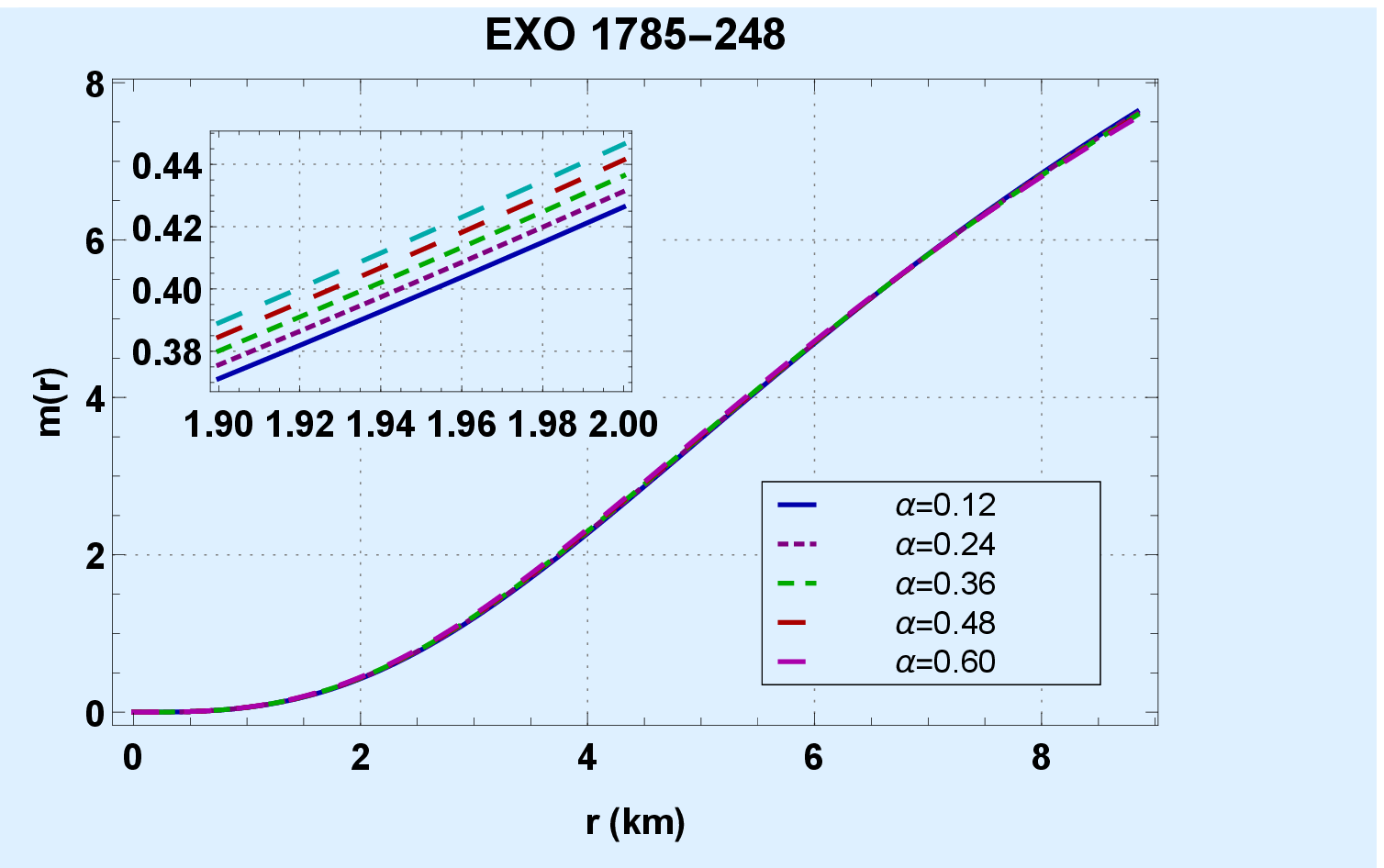}
        \includegraphics[scale=.37]{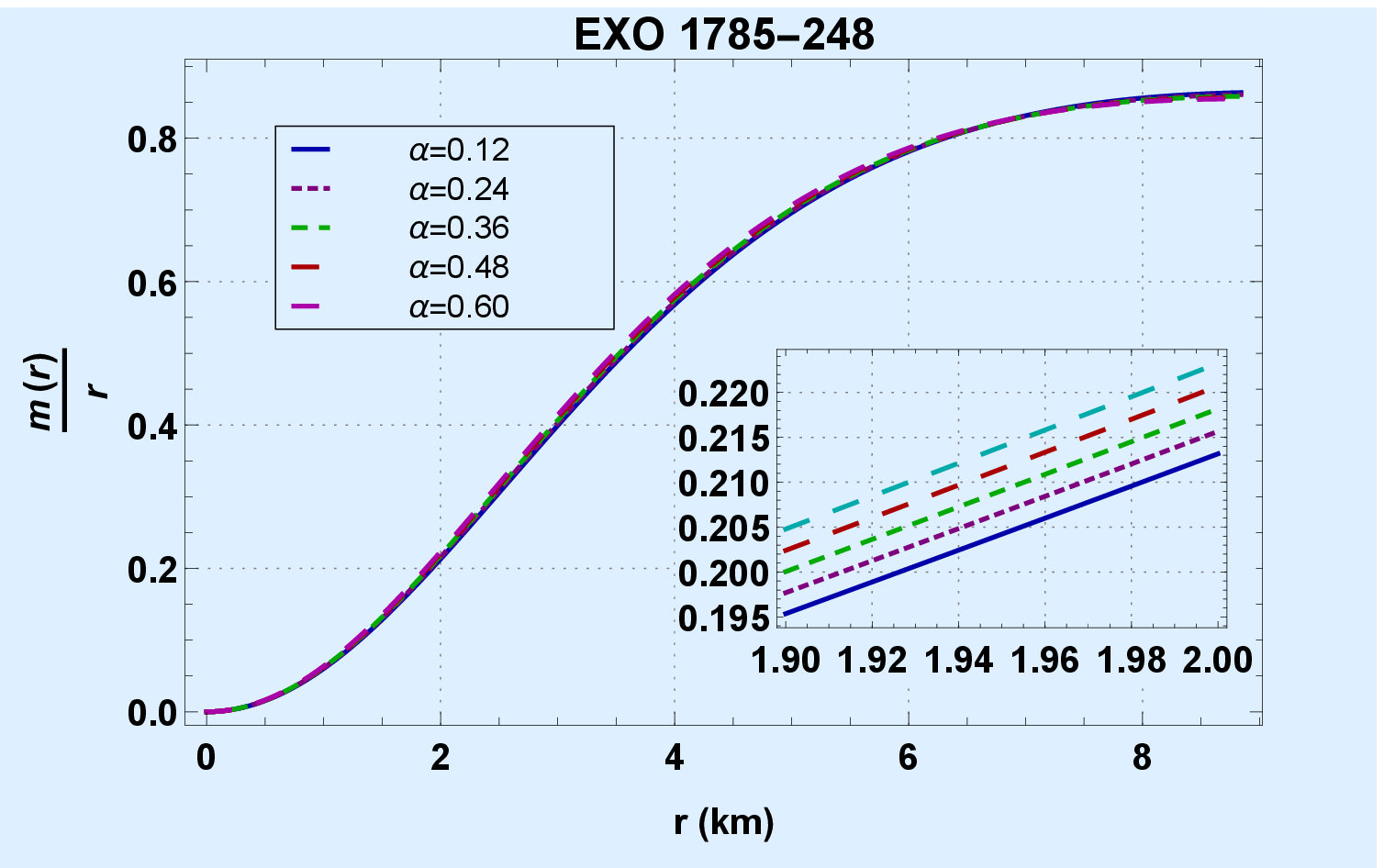}
        \includegraphics[scale=.45]{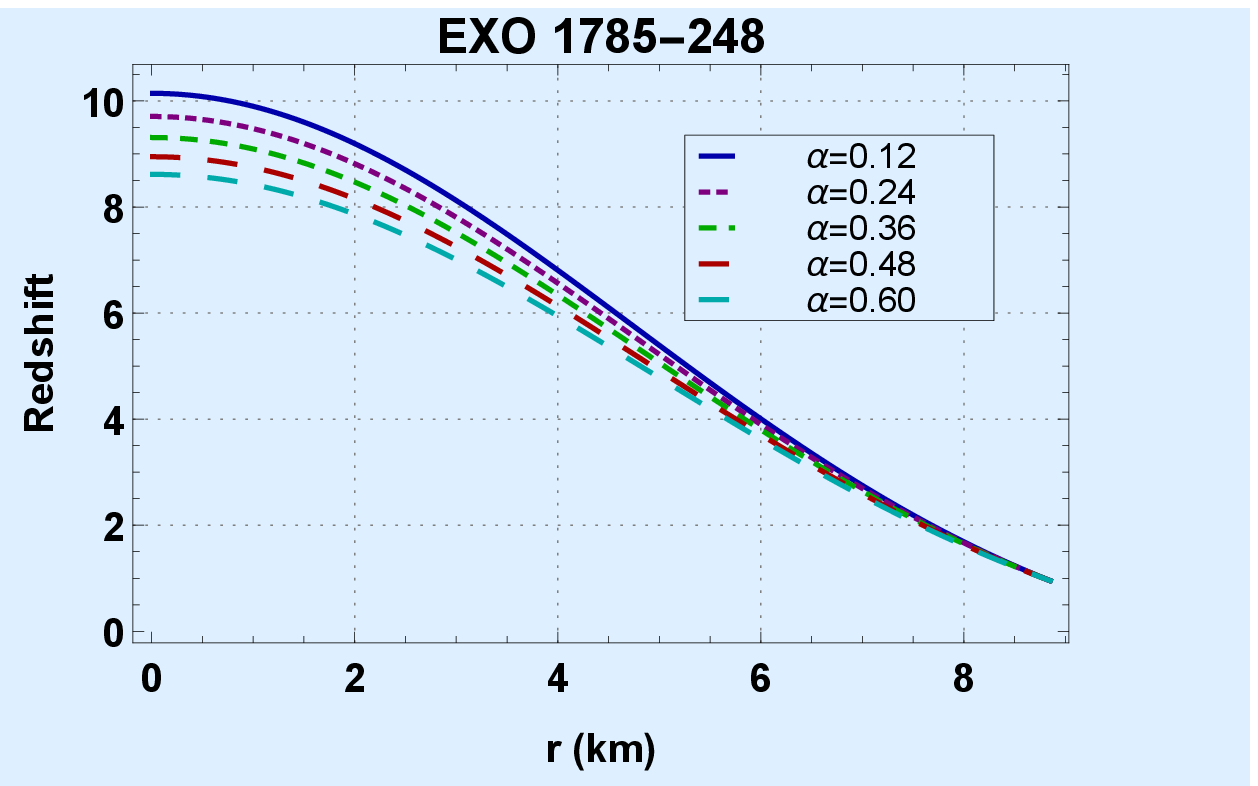}
        \includegraphics[scale=.37]{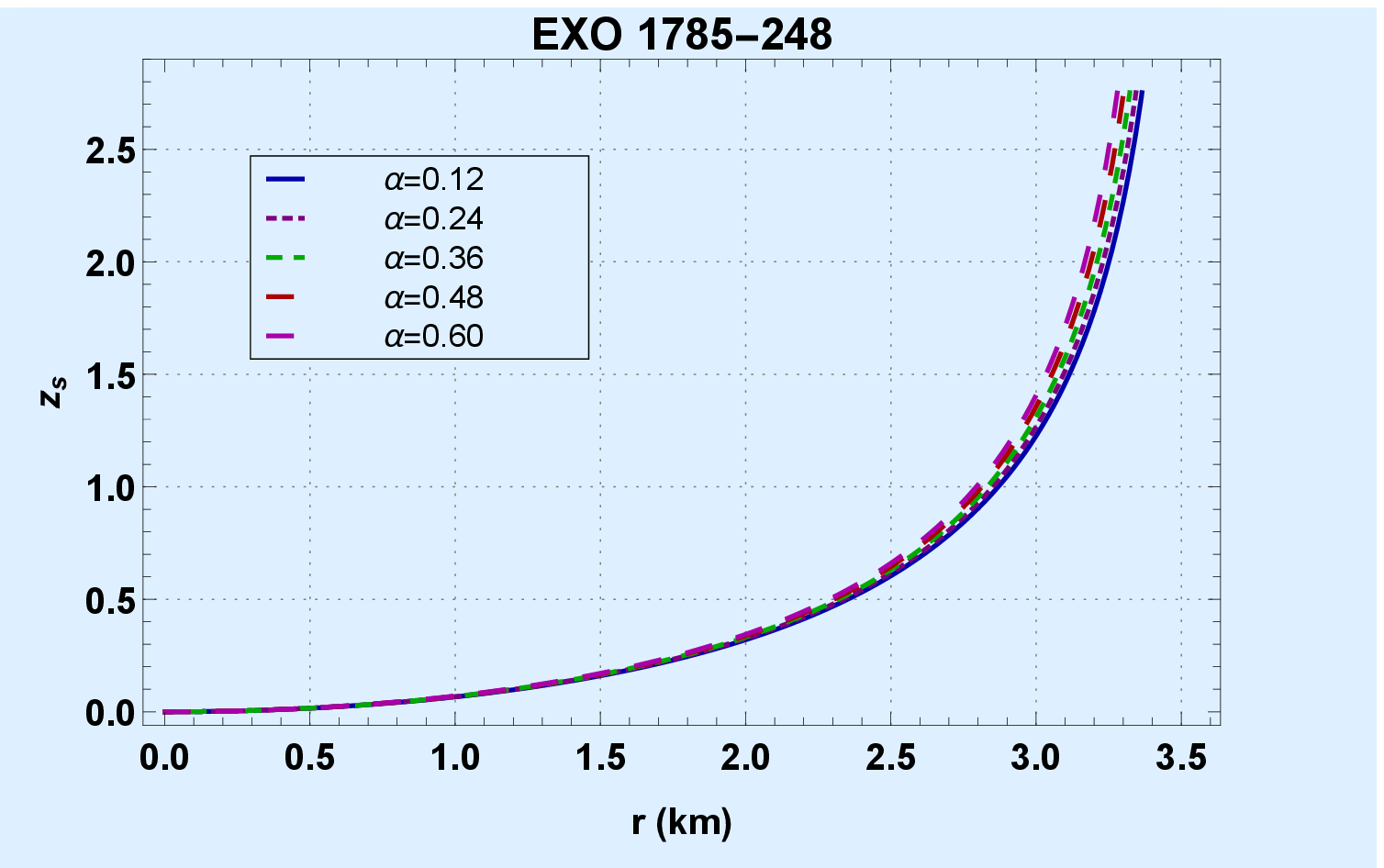}
        \caption{Variation of (I) Mass function, (II) compactness factor, (III) gravitational redshift, and (IV) surface redshift versus radial coordinate $r$ for the compact object for different values of $\alpha$.\label{mass}}
\end{figure}

These four quantities' behaviors, namely, mass function, compactness factor, gravitational redshift, and surface redshift, are displayed in Fig. \ref{mass}. The graph demonstrates that each physically significant variable meets the criteria of a realizable stellar configuration. It should be noted that the mass function at the center is guaranteed to be regular. We can also see that the three quantities, namely mass function, compactness factor, and surface redshift are monotonically increasing functions, but the quantity: gravitational redshift is a monotonically decreasing function as a function of the radius $r$ and also all these quantities are positive inside the stellar body. In particular, the plot demonstrates that the mass-radius ratio and its lower and upper limits for the compact stellar candidate under consideration in this research comply completely with the criteria listed in Table \ref{table1}, which may be confirmed in \cite{Bohmer06}. However, the authors \cite{Ivanov02,Bohmer06} contend that the surface redshift of an anisotropic fluid sphere is hypothesized to be less than $z_s\leq 5$ or $z_s\leq 5.211$. Hence, under these restrictions, the surface redshift is fulfilled everywhere within the stellar configuration, which demonstrates the viability of our stellar model.

\begin{figure}[H]
    \centering
        \includegraphics[scale=.465]{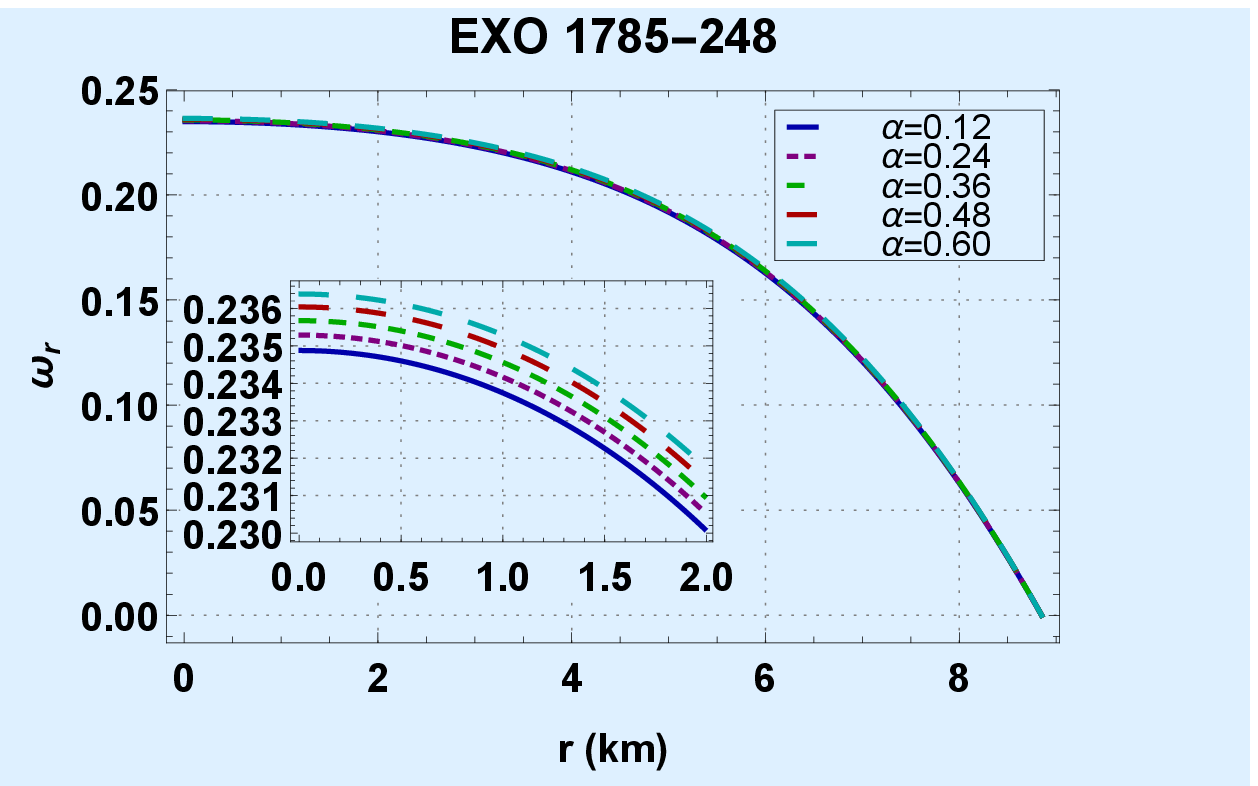}
        \includegraphics[scale=.465]{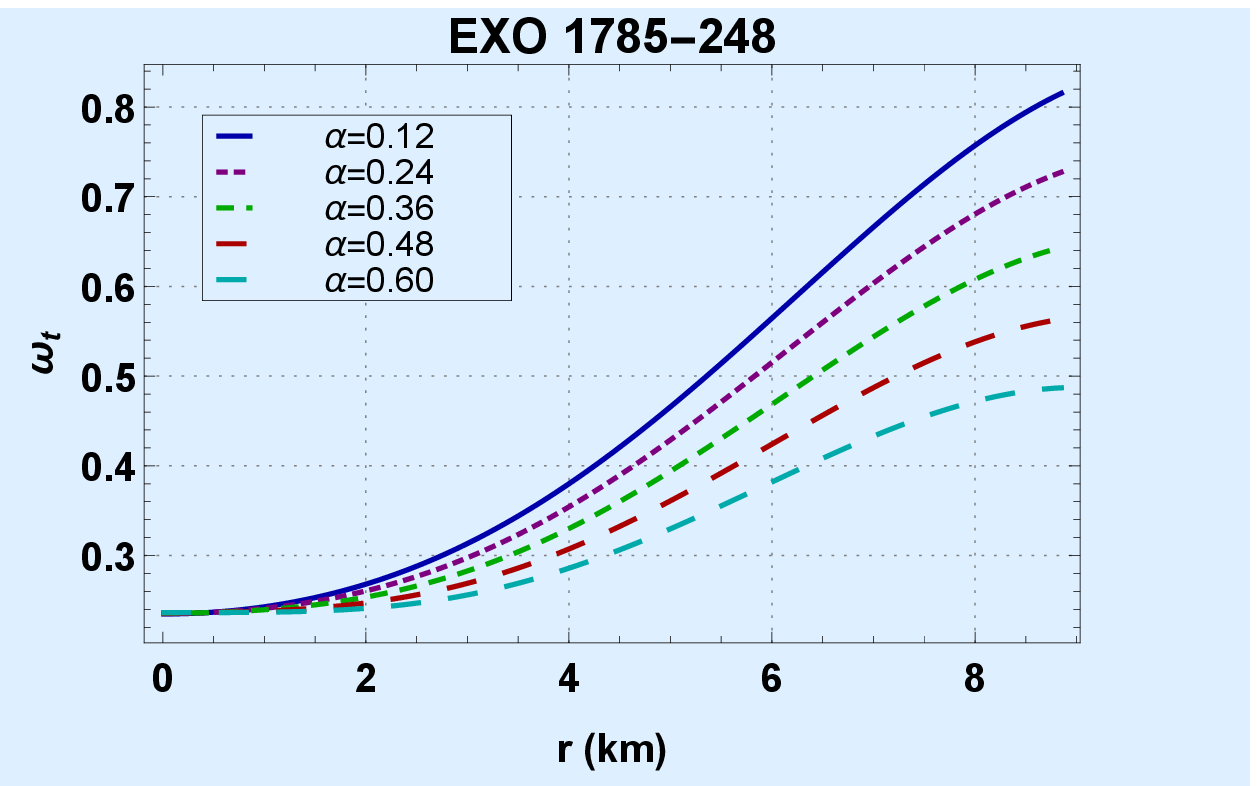}
        \includegraphics[scale=.46]{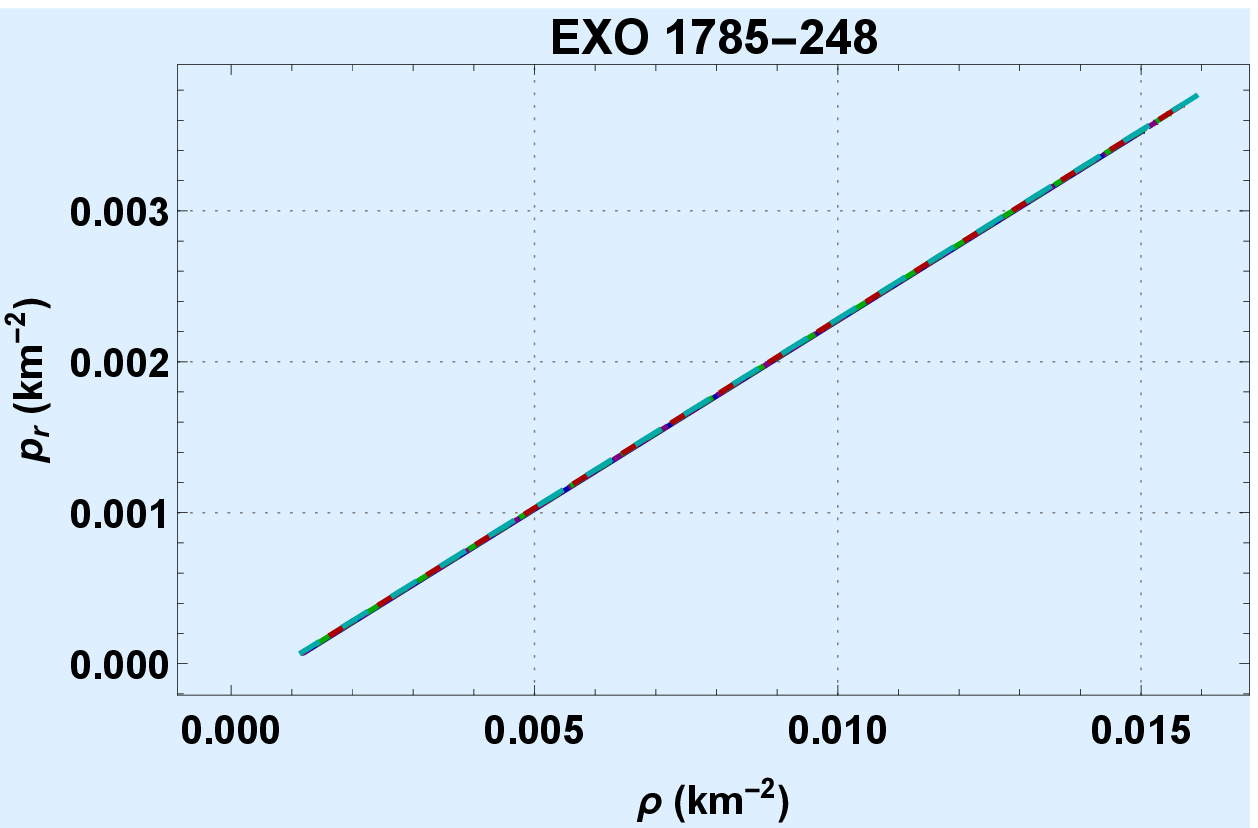}
        \includegraphics[scale=.465]{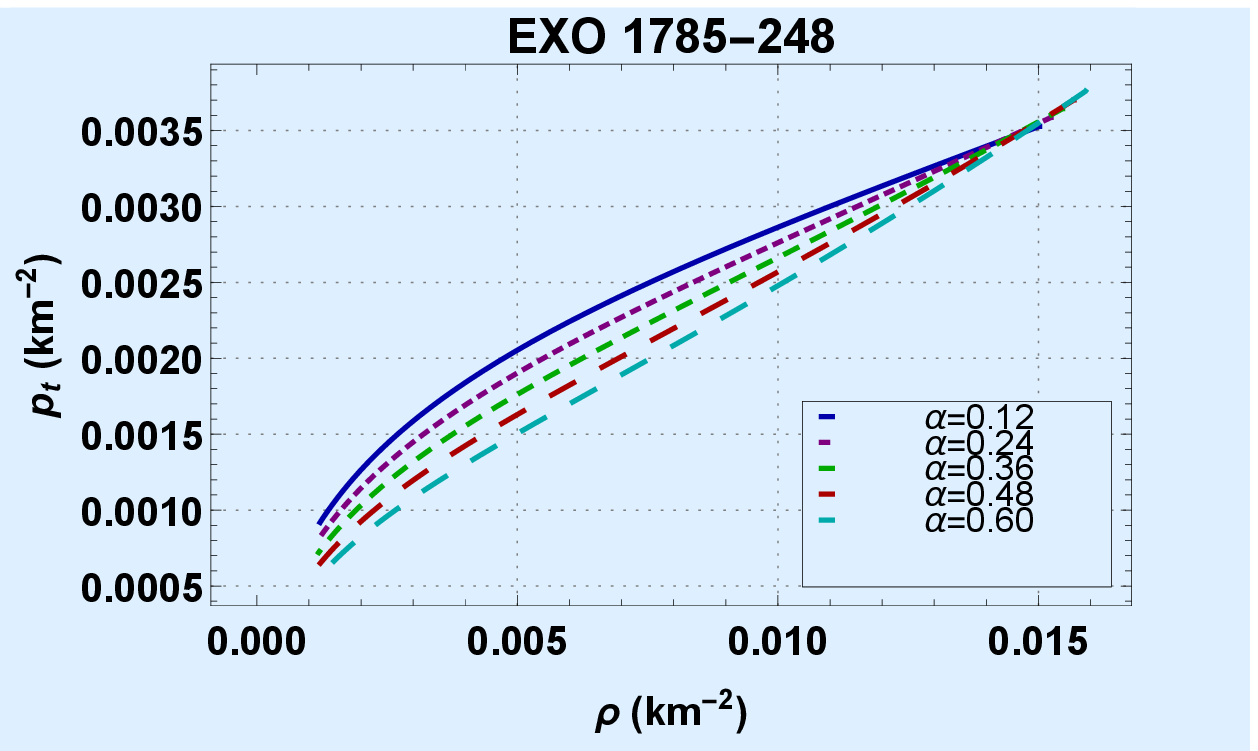}
       \caption{(Left) EoS parameter $\omega_r$ is shown against the radial distance $r$, (Middle) EoS parameter $\omega_t$ is shown against the radial distance $r$, (Right) Relation between $p_r$ and $\rho$, (Bottom) Relation between $p_t$ and $\rho$.  }
    \label{eoss}
\end{figure}

Moreover, we analyzed whether the charged matter is realistic or exotic based on various constraints applicable to EoS components such as $0\leq w_r<1,\;\;\&\;\;0<w_t<1$. These matter-related constraints should be maintained for a realistically charged and uncharged matter composition. The EoS has the following mathematical forms,
\begin{equation}\label{58}
   w_r = \frac{p_r}{\rho}\;\;\;\;\& \;\;\;\;w_t = \frac{p_t}{\rho}.
\end{equation}
It is straightforward to establish that our model of the heavenly objects faithfully reflects the realistic nature of the composition of matter under the radial and transverse components of the EoS. These $w_r,\;\; \&\;\;w_t$  profiles' graphs are displayed in Fig. \ref{eoss}. The evolution of radial and transverse pressures versus density is also shown in Fig. \ref{eoss} for different values of $\alpha$.

\begin{table*}[hbt!]
\centering
\caption{{ Comparative study of lower bound, mass-radius ratio, compactness, upper bound, and surface red-shift of the
stellar toy model for different values of $\alpha$ }}\label{table1}
 \scalebox{0.95}{\begin{tabular}{|c|c|c|c|c|c|c|c|c|c|c|}\hline
{$\alpha$}  & Lower bound &  Mass-radius & compactness ($u$)   & Upper bound & Surface redshift\\
 & $\frac{Q^2\,(18R^2+Q^2)}{2R^2\,(12R^2+Q^2)}$ & ratio ($\frac{{M}}{R}$)& $\frac{{M}_{\text{eff}}}{R}$ &$\frac{2R^2+3Q^2+2R\sqrt{R^2+3Q^2}}{9R^2}$& $z_s$\\ \hline

 0.12  &  0.0215284  & 0.383426 & 0.369062 & 0.463398 &  0.954124  \\\hline

 0.24   &  0.0215284  & 0.383424 & 0.369061 & 0.463398 &  0.954115 \\\hline

 0.36  &  0.0215284  & 0.383423 & 0.36906 & 0.463398 &  0.954106  \\\hline

 0.48  &  0.0215284  & 0.383422 & 0.369058 & 0.463398 &  0.954097  \\\hline

 0.60  &  0.0215284  & 0.383421 & 0.369057 & 0.463398 &  0.954088  \\\hline
\end{tabular}}
\end{table*}
\section{Concluding remarks}\label{Sec8}
In this paper, we have extensively explored the viability and stability of a static charged anisotropic fluid sphere consisting of a charged perfect fluid in the framework of 5D EMGB gravity theory. What's more, we have employed the well-behaved TK {\it ansatz} together with a linear EoS of the form $p_r=\beta \rho-\gamma$, (where $\beta$ and $\gamma$ are constants) to generate exact solutions of the EMGB field equations and analyze the effect of the Gauss-Bonnet Lagrangian term $\mathcal{L}_{GB}$ which is coupled with the Einstein-Hilbert action through the coupling constant $\alpha$ on the main physical properties of the stellar model.

The eventual model has been thoroughly analyzed to identify whether it conforms to strict regularity and stability conditions. In addition, the causality condition via Herrera's cracking concept and the relativistic adiabatic index was deemed to be satisfactory. The adherence of the principal features such as energy density, pressure components, anisotropy factor, sound velocity, hydrostatic equilibrium of the stellar system via a modified TOV equation, mass-radius relation, compactness factor, surface redshift with the recognized physical behavior for different values of coupling constant $\alpha$ has been confirmed by graphical plots using the stated parametric values. Furthermore, our analysis demonstrates that matter configuration attributes and higher dimensional effects are intimately associated. The interior spacetime and the exterior space-time described by the EGB Schwarzschild metric were also matched to settle all integration constants that arose along the way.

This study has shown the meaning and significance of the Gauss-Bonnet Lagrangian term $\mathcal{L}_{GB}$ which is coupled with the Einstein-Hilbert action through the coupling constant $\alpha$ together with the TK {\it ansatz} and a linear EoS in the development of astrophysical models that harmonize with observed data.

\section*{Acknowledgments}
AE thanks the National Research Foundation (NRF) of South Africa for the award of a postdoctoral fellowship.
\section*{Declarations}
\textbf{Data Availability Statement:} The results are obtained
via purely theoretical calculations and can be verified analytically,
thus this manuscript has no associated data, or the data will not be deposited. \par
\textbf{Conflicts of Interest:} The authors have no financial interest or involvement which is relevant by any means to the content of this study.


\begin{thebibliography}{99}
\bibitem{Riess-Perlmutter:1998} 
A. G. Riess et al., Astron. J., 116, 1009 (1998);
\bibitem{Riess-Perlmutter:1999}S. Perlmutter et al., Astrophys. J., 517 565 (1999).
\bibitem{de-Rham:2014} 
C. de Rham. Living Rev. Rel. 17, 7 (2014).
\bibitem{Horndeski:1974}
G. W. Horndeski, Int. J. Theor. Phys., 10, 363 (1974).
\bibitem{Gross:1999}
D. Gross, Nucl. Phys. Proc. Suppl., 74, 426 (1999).
\bibitem{Hansraj:2022}
S. Hansraj, Eur. Phys. J. C 82, 218 (2022).
\bibitem{Mustafa:2022}
G. Mustafa, A. Errehymy, A. Ditta, M. Daoud, Chin. J. Phys. 77, 2781-2794 (2022).

\bibitem{Ghosh:2014}
S. G. Ghosh, S. Jhingan and D. W. Deshkar, J. Phys.: Conf. Series, 484, 012013 (2014).
\bibitem{tol} R.C. Tolman, Phys. Rev. 55, 364 (1939).
\bibitem{kuch} B. Kuchowicz, Acta Phys. Pol. 33, 541 (1968).
\bibitem{Kaluza:1921}
T. Kaluza, 1921. Sitz. Ber. Preuss. Akad. Wiss., 966 (1921).
\bibitem{Klein:1926}
O. Klein, Zeit. f. Physik , 37, 895 (1926).

\bibitem{Tangherlini:1963}
F. R. Tangherlini, Il Nuovo Cimento, 27, 636 (1963).
\bibitem{Myers:1986}
R. C. Myers and M. J. Perry, Ann. Phys., 172, 304 (1986).
\bibitem{Myers:1988}
R. C. Myers and J. Z. Simons, Phys. Rev. D, 38, 2434 (1988).
\bibitem{Cai:2002}
R.G. Cai, Phys. Rev. D 65, 084014 (2002).
\bibitem{Cai:2004}
R.G. Cai, Q. Guo, Phys. Rev. D 69, 104025 (2004).
\bibitem{Giacomini:2015}
A. Giacomini, J. Oliva, A. Vera, Phys. Rev. D 91, 104033 (2015).
\bibitem{Xu:2015}
W. Xu, J. Wang, X. h. Meng, Phys. Lett. B 742, 225 (2015).
\bibitem{Aranguiz:2016}
L. Aranguiz, X.M. Kuang, O. Miskovic, Phys. Rev. D 93, 064039
(2016)
\bibitem{Ghosh:2017}
S.G. Ghosh, M. Amir, S.D. Maharaj, Eur. Phys. J. C 77, 530 (2017).

\bibitem{Bhawal:1990}
B. Bhawal, Phys. Rev. D 42, 449 (1990)
\bibitem{Gallo:2015}
E. Gallo, J.R. Villanueva, Phys. Rev. D 92, 064048 (2015)
\bibitem{Wu:2021}
C.H. Wu, Y.P. Hu, H. Xu, Eur. Phys. J. C 81, 351 (2021)
\bibitem{Xu:2019}
W. Xu, C. y Wang, B. Zhu, Phys. Rev. D 99, 044010 (2019)
\bibitem{Ghosh:2018}
S.G. Ghosh, D.V. Singh, S.D. Maharaj, Phys. Rev. D 97, 104050 (2018)
\bibitem{Maeda:2008}
H. Maeda, M. Nozawa, Phys. Rev. D 78, 024005 (2008)
\bibitem{Mehdizadeh:2015}
M. R. Mehdizadeh, M. Kord Zangeneh, F. S. N. Lobo, Phys. Rev. D 91, 084004 (2015).
\bibitem{Jhingan:2010}
S. Jhingan, S.G. Ghosh, Phys. Rev. D 81, 024010 (2010).
\bibitem{Maeda:2006}
H. Maeda, Phys. Rev. D 73, 104004 (2006).
\bibitem{Zhou:2015}
K. Zhou, Z.Y. Yang, D.C. Zou, R.H. Yue, Mod. Phys. Lett. A 26, 2135 (2015).
\bibitem{Abbas:2015}
G. Abbas, M. Zubair, Mod. Phys. Lett. A 30, 1550038 (2015).

\bibitem{Demorest:2010}
P. Demorest, T. Pennucci, S. Ransom, M. Roberts, J. Hessels, Nature 467, 1081 (2010).
\bibitem{Antoniadis:2013}
J. Antoniadis , P.C.C. Freire, N. Wex et al., Science 340, 6131 (2013).
\bibitem{Lake:2003}
K. Lake, Phys. Rev. D 67, 104015 (2003).
\bibitem{Herrera:2008}
L. Herrera, J. Ospino, A. Di Prisco, Phys. Rev. D 77, 027502 (2008).
\bibitem{Schwarzschild:1916}
K. Schwarzschild, Sitzungsber. Preuss. Akad. Wiss. Berlin (Math. Phys. ) 1916, 189 (1916).
\bibitem{Tolman:1939}
R.C. Tolman, Phys. Rev. 55, 364 (1939).

\bibitem{refa0} S. Capozziello and M. De Laurentis, Phys. Rept. 509,
167-321 (2011)
\bibitem{refa1}S. Capozziello, A. Finch, J. L. Said, and A. Magro, Eur. Phys. J. C, 81, 12, 1141, (2021)
\bibitem{refa2}K. F. Dialektopoulos, T. S. Koivisto, and S. Capozziello, Eur. Phys. J. C 79, 606 (2019).

\bibitem{refb0} R. Goswami,  Nzioki A, Maharaj S. D and  Ghosh S. G,  Phys. Rev. D 90, 084011 (2014).
\bibitem{refb1}Chilambwe B., S.  Hansraj and S. D. Maharaj, Int. J. Mod. Phys. D, 24 1550051 (2015).
\bibitem{refb2} M. Govender,  A. Maharaj ,  D. Lortan and  D.Day, Astrophys. Space. Sci., 363, 165 (2018).
\bibitem{refb3} Govender M.,  N. Mewalal and  S. Hansraj, Eur. Phys. J. C, 79, 24 (2019).
\bibitem{refb4} S. Hansraj,  Chilambwe B, Maharaj, Eur. Phys. J. C, 27 277 (2015).
\bibitem{refb5}S. Hansraj,  Govender M.,  Banerjee A. and  Mkhize N., Classical and Quantum Gravity, 38, 065018 (2021).

\bibitem{refc0}
S. K. Maurya, A. Errehymy, R. Nag and M. Daoud,
Fortsch. Phys. 70 2200041 (2022).
\bibitem{refc1}S. K. Maurya, A. Errehymy, M. K. Jasim et al.,  Eur. Phys. J. C, 82, 1173 (2022).
\bibitem{refc2} A. Errehymy G. Mustafa, Y Khedif et al., Chin. Phys. C, 46, 045104 (2022).
\bibitem{refc3}A. Errehymy, Y. Khedif, G. Mustafa, et al., Chin. J. Phys., 77, 1502-1522 (2022).
\bibitem{refc4}A. Ditta, A. Errehymy, X. Tiecheng, et al., Eur. Phys. J. Plus, 137, 933 (2022).
\bibitem{refc5} A. Errehymy, A. Ditta, G. Mustafa et al., Eur. Phys. J. Plus, 137, 1311 (2022). 
\bibitem{refd0} T. Tangphati, A. Pradhan, A. Errehymy, A. Banerjee, Phys. Lett. B, 819, 136423 (2021).
\bibitem{refd1} Tangphati, T., A. Pradhan, A. Errehymy, A. Banerjee, Ann. Phys., 430, 168498 (2021).
\bibitem{refd2} A. Errehymy, Y. Khedif and M. Daoud, Eur. Phys. J. C 81, 266 (2021).
\bibitem{refd3}S.K. Maurya, K. N. Singh, M. Govender, et al., Eur. Phys. J. C, 81, 729 (2021).
\bibitem{refd4} A. Errehymy, Y. Khedif, M. Daoud, Eur. Phys. J. C, 81, 266 (2021).

\bibitem{refe0}S.K. Maurya, A. Errehymy, D. Deb, et al., Phys. Rev. D. 100, 044014 (2019).
\bibitem{refe1}F. Tello-Ortiz, S.K. Maurya, A. Errehymy, et al., Eur. Phys. J. C 79, 885 (2019).

\bibitem{reff0}S. K. Maurya et al., Phys. Dark Univ., 30100620 (2020).
\bibitem{reff1}K. N. Singh et al., Chin. Phys. C, 44 105106 (2020).
\bibitem{reff2} M. Rahaman,  K.N. Singh, A. Errehymy et al., Eur. Phys. J. C 80, 272 (2020).
\bibitem{reff3} S.K. Maurya, K. N. Singh, A. Errehymy, et al., Eur. Phys. J. Plus 135, 824 (2020).
\bibitem{reff4}K. N. Singh, A. Errehymy, F. Rahaman al., Chinese Phys. C 44 105106 (2020).
\bibitem{reff5}K. N. Singh, S. K. Maurya, A. Errehymy, et al., Phys. Dark Universe, 30, 100620 (2020).
\bibitem{Boulware}D.G. Boulware, S. Deser, Phys. Rev. Lett. 55, 2656 (1985).
\bibitem{Lake:2003}
K. Lake, Phys. Rev. D, 67, 104015 (2003).
\bibitem{her1} Herrera, L.  Phys. Lett. A 1992, 165, 206-210.
\bibitem{her2}DiPrisco, A., Fuenmayor, E., Herrera, L., Varela, V. Phys. Lett. A 1994, 195, 23-26.
\bibitem{her3}DiPrisco, A.; Herrera, L.; Varela, V. Gen. Rel. Grav. 1997, 29, 1239.
\bibitem{abr}H. Abreu, H. Hernandez, L.A. Nunez, Class. Quantum
Grav. 24, 4631 (2007).

\bibitem{Bohmer06} G. B\"ohmer, T. Harko, Class. Quantum Gravity 23, 6479 (2006).
\bibitem{Ivanov02} B.V. Ivanov, Phys. Rev. D 65, 104001 (2002). 
\bibitem{chandra} S. Chandrasekhar, Astrophys J. 140, 417 (1964)
\bibitem{ad1} W. Hillebrandt, K.O. Steinmetz, Astron. Astrophys. 53, 283 (1976)
\bibitem{ad2} D. Horvat, S. Ilijic, A. Marunovic, Class. Quantum Gravity 28, 025009 (2011)
\bibitem{ad3} D.D. Doneva, S.S. Yazadjiev, Phys. Rev. D 85, 124023 (2012)
\bibitem{ad4} H.O. Silva, C.F.B. Macedo, E. Berti, L.C.B. Crispino, Class. Quantum Gravity 32, 145008 (2015)

\end{thebibliography}
\end{document}